\documentclass[journal]{IEEEtran}
\usepackage[OT1]{fontenc} 
\usepackage{cite}
\usepackage{float}
\usepackage{multicol,multirow}
\usepackage{makecell,booktabs}
\usepackage[colorlinks=true,linkcolor=black,anchorcolor=black,citecolor=black,filecolor=black,menucolor=black,runcolor=black,urlcolor=black]{hyperref}

\usepackage{graphicx}
\usepackage{amsfonts}
\usepackage{amssymb}
\usepackage{amsmath}
\usepackage{amsthm}
\usepackage{mathtools}
\usepackage{array}
\newcolumntype{P}[1]{>{\centering\arraybackslash}p{#1}}
\usepackage{color}
\usepackage{epstopdf}
\usepackage{subcaption}
\usepackage{bbm}
\usepackage{bm}
\usepackage{comment}
\usepackage{forest}
\usepackage{nomencl}
\usepackage{dblfloatfix}
\usepackage{enumitem,kantlipsum}

\usepackage{etoolbox}
\usepackage[noend]{algpseudocode}
\makeatletter
\def\BState{\State\hskip-\ALG@thistlm}
\makeatother

\usepackage[ruled]{algorithm2e}
\SetKwRepeat{Do}{do}{while}

\newcommand{\norm}[1]{\left\lVert#1\right\rVert}

\DeclareMathOperator{\Stdev}{Stdev}
\DeclareMathOperator{\Norm}{Norm}

\begin{document}
    \title{Computing a Strategic Decarbonization Pathway: \\A Chance-Constrained Equilibrium Problem}
    \author{Jip~Kim,~\IEEEmembership{Student Member,~IEEE,}~Robert~Mieth,~\IEEEmembership{Student Member,~IEEE,} and~Yury~Dvorkin,~\IEEEmembership{Member,~IEEE\vspace{-4mm}}%
    }
    
    \maketitle    
    \begin{abstract}
        US transmission systems and wholesale electricity markets, albeit federally regulated, often span across multiple state jurisdictions. In this environment, state regulators can strategically exploit this techno-economic coupling to  advance their  clean energy policy goals at the expense of neighboring jurisdictions. This paper investigates strategic regulatory competition to understand its effect on achieving Renewable Portfolio Standards (RPS). We formulate a chance-constrained equilibrium problem with equilibrium constraints (CC-EPEC), which considers multiple state regulators, acting in coordination with in-state power companies,  to implement RPS goals in the least-cost manner. To solve this CC-EPEC, we customize a Progressive Hedging (PH) algorithm. The case study uses the CC-EPEC and PH algorithm to analyze the effects of state regulatory competition in the ISO New England system.        
    \end{abstract}

\vspace{-2mm}
\section*{Nomenclature}
\vspace{-3mm}
\subsection{Sets and Indices}
\addcontentsline{toc}{subsection}{Sets and Indices}
\begin{IEEEdescription}[\IEEEusemathlabelsep\IEEEsetlabelwidth{$n\in {\cal{N}}^{\mathrm{T/D}}$}]
\item[$e\in {\cal{E}}$]{Set of representative operating days}
\item[$i\in {\cal{I}}$]{Set of existing generators}
\item[$i\in \hat{\cal{I}}$]{Set of candidate generators}
\item[$l\in {\cal{L}}$]{Set of transmission lines}
\item[$n\in {\cal{N}}$]{Set of transmission nodes}
\item[$s\in {\cal{S}}$]{Set of states, ${\cal{S}}\!\!=$\{ME, NH, VT, MA, CT, RI\}}
\item[$t\in {\cal{T}}$]{Set of time intervals}
\item[${\cal{I}^{\mathrm{R}}}, {\cal{I}^{\mathrm{C}}} \subset {\cal{I}}$]{Set of renewable/controllable generators}
\item[${\cal{I}}_n \subset {\cal{I}}$]{Set of generators located on bus $n$}
\item[${\cal{I}}_s \subset {\cal{I}}$]{Set of generators of regulator $s$}
\item[${\cal{N}}_s \subset {\cal{N}}$]{Set of transmission nodes of regulator $s$}
\item[$\text{SOS1}$]{Special ordered set of type 1}
\item[${\mathcal{K}}^n$]{Second order conic set}
\item[${r(l), o(l)}$]{Receiving/sending node of line $l$}
\item[${n(i)}$]{Node where generator $i$ is located}
\item[${s(i)}$]{State where generator $i$ is located}
\end{IEEEdescription}

\vspace{-3mm}
\subsection{Parameters}
\addcontentsline{toc}{subsection}{Parameters}
\begin{IEEEdescription}[\IEEEusemathlabelsep\IEEEsetlabelwidth{${P}_n^{\downarrow,\mathrm{max}}\!\!\!$}]
\item[$\alpha_{it}$]{Participation parameter of controllable generator $i$}
\item[$\bm{\epsilon}_{ite}$]{Forecast error of renewable generator $i$}
\item[${\varepsilon}$]{Progressive hedging termination tolerance}
\item[$\rho_{ite}$]{Forecast factor of generator $i$ on time $t$}
\item[$\rho^{\mathrm{g}}, \rho^{\lambda}$]{Progressive hedging penalty associated with $g, \lambda$}
\item[$\kappa_{s}$]{Renewable portfolio standard goal of regulator $s$}
\item[$\pi^{\mathrm{D}}_{nt}$]{Retail electricity tariff of zone $n$}
\item[$\sigma_{ite}$]{Normalized standard deviation of generation forecast error of unit  $i$ [$\mathrm{MW}$]}
\item[$\upsilon_{ite}$]{Normalized mean of generation forecast error of unit $i$ [$\mathrm{MW}$]}
\item[$\omega_{e}$]{Probability of representative day $e$, ($\sum{\omega_e}=1$)}
\item[$\Delta\lambda$]{Size of a discretization step for the binary expansion approach [$\$/\mathrm{MW}$]}
\item[$\Gamma_i$]{Minimum output limit factor of controllable generator $i$}
\item[$P^{\mathrm{CT}}_{s}$]{Capacity tariff of regulator $s$ [$\$/\mathrm{kW}$]}
\item[$P^{\mathrm{ET}}_{s}$]{Energy feed-in-tariff of regulator $s$ [$\$/\mathrm{MWh}$]}
\item[$B^{\mathrm{C}}_s$]{Budget for generation expansion of regulator $s$}
\item[$B^{\mathrm{P}}_s$]{Budget for renewable policies of regulator $s$}
\item[$C^{\mathrm{inv}}_{i}$]{Capital cost of generator $i$ (pro-rated on a daily basis using the net present value approach) [$\mathrm{\$/MW}$]}
\item[$C^{\mathrm{g}}_{i}$]{Incremental cost of generator $i$ [$\mathrm{\$/MWh}$]}
\item[$D_{nte}$]{Real power demand of zone $n$ [$\mathrm{MW}$]}
\item[${F}^{\mathrm{max}}_l$]{Apparent flow limit of line $l$ [$\mathrm{MVA}$]}
\item[${G}^{\mathrm{min}}_{it}$]{Lower bound of generation power output of existing generator $i$ [$\mathrm{MW}$]}
\item[${G}^{\mathrm{max}}_{it}$]{Upper bound of generation power output of existing generator $i$ [$\mathrm{MW}$]}
\item[${H}^{\mathrm{max}}_i$]{Upward ramping limit of generator $i$ [$\mathrm{MW/h}$]}
\item[${H}^{\mathrm{min}}_i$]{Downward ramping limit of generator $i$ [$\mathrm{MW/h}$]}
\item[${K}$]{Discretization parameter for the binary expansion approach}
\item[${P}_n^{\downarrow,\mathrm{max}}$]{Apparent flow limit of the interface line into region $n$ from the transmission network [$\mathrm{MVA}$]}
\item[${R}_{ite}$]{Allocated reserve capacity of generator  $i$ [$\mathrm{MW}$]}
\item[$X_l$]{Reactance of transmission line $l$ [$\Omega$]}
\end{IEEEdescription}

\vspace{-3mm}
\subsection{Variables}
\addcontentsline{toc}{subsection}{Variables}
\begin{IEEEdescription}[\IEEEusemathlabelsep\IEEEsetlabelwidth{$m^{(i)}_s,,,$}]
\item[$f^{\mathrm{p}}_{lte},f^{\mathrm{q}}_{lte}$]{Real/Reactive power flow of line $l$ [$\mathrm{MW,MVar}$]}
\item[$\bm{g}_{ite}$]{Real power output of generator $i$ [$\mathrm{MW}$]}
\item[$\overline{g}_{ite}$]{Expected real power output of generator $i$ [$\mathrm{MW}$]}
\item[${g}^{\mathrm{o}}_{ite}$]{Offer of generator $i$ into wholesale market  [$\mathrm{MW}$]}
\item[${g}_{ite}$]{Dispatch signal of the wholesale market for generator $i$ [$\mathrm{MW}$]}
\item[$g^{\mathrm{max}}_{i}$]{Investment decision variable of generator $i$ [$\mathrm{MW}$]}
\item[$m^{(i)}_s\!,w^{(i)}_s$]{Progressive hedging multiplier associated with $g,\lambda$}
\item[$p^{\downarrow}_{nte}$]{Interface power flow into region $n$ from the transmission network [$\mathrm{MW}$]}
\item[$\lambda_{nte}$]{Locational marginal price in region $n$ [$\$/\mathrm{MW}$]}
\item[$\theta_{nte}$]{Node angle of transmission node $n$}
\end{IEEEdescription}

\vspace{-3mm}
\subsection{Operators}
\addcontentsline{toc}{subsection}{Operators}
\begin{IEEEdescription}[\IEEEusemathlabelsep\IEEEsetlabelwidth{${\Stdev[\cdot]}$}]
\item[${\norm{\cdot}_2}$]{Euclidean norm (2-norm)}
\item[${\Stdev[\cdot]}$]{Standard deviation}
\item[${\mathrm{card}}(\cdot)$]{Cardinality of a set}
\end{IEEEdescription}
{\normalsize Random variables/parameters are in \textbf{bold} fonts}

\section{Introduction}
Motivated by the mitigation of climate change, decarbonization policies have gained a lot of attention around the world. In the U.S. electric power sector, state regulators lead such policy efforts by setting the renewable portfolio standards (RPS) in their jurisdiction. RPS goals vary, but typically aim to achieve a certain amount of electricity supply from renewable energy resources (RES) by a target year. To support these goals, there are various incentives that aim to shift power production from fossil-fueled generation resources to RES, including such policy measures as subsidies, tax credits, carbon tax, etc. \cite{irena2015renewable,ec2013support}. However, RPS are energy-centric and do not explicitly account for an increased need in the provision of ancillary services to support a major increase in RES productions. Since RES are limited in their ability to provide ancillary services (in particular, active power reserves), anticipated retirement of coal or nuclear resources may reduce the number of generators  capable of providing reserve, thus obstructing further RES integration and implementation of RPS goals. 

\begin{figure}[t]
    \vspace{-0mm}
    \centering    
    \includegraphics[width=0.76\columnwidth]{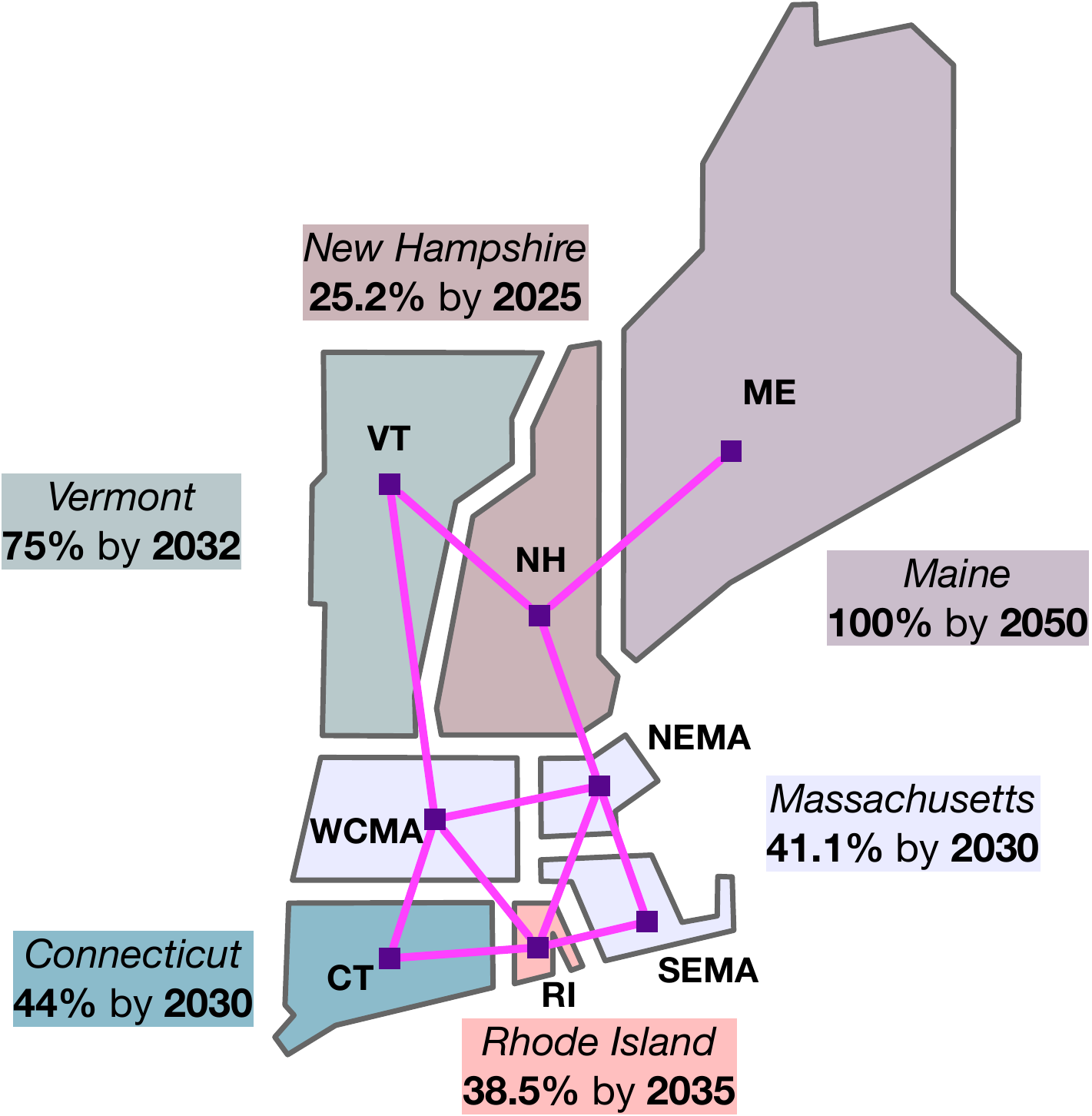}
    \caption{\small ISO New England 8-zone test system with renewable portfolio standard goals as of January 2020. \vspace{-3mm}}
    \label{fig:platform}   
    \vspace{-0mm}
\end{figure} 

Historically, expansion of grid infrastructure has been planned in a centralized framework, i.e. with a single entity that makes operation and expansion decisions in the entire system. Munoz \textit{et al.}\cite{munoz2014engineering} investigated the effect of different renewable policy scenarios on  expansion decisions under this framework. In a deregulated electricity market, centralized planning with system-wide policy scenarios as in \cite{munoz2014engineering} may yield inefficient results as power grids and wholesale markets often span multiple jurisdictional boundaries. For example, the ISO New England transmission system and market, which are federally regulated, cover six state jurisdictions with different RPS goals as illustrated in Fig.~\ref{fig:platform} \cite{krishnamurthy20168}. In this environment, this techno-economic coupling can be exploited by strategically acting state regulators to advance their policy goals at the expense of other states. Therefore, this strategic behavior should be analyzed to properly assess the effectiveness of renewable policies and power grid expansion. 

Previously, strategic behavior in electricity markets has been  investigated in the context of  expansion planning, e.g.\cite{wogrin2011generation,baringo2013strategic,jalal2015investment}, to analyze strategic interactions between a market participant and the wholesale electricity market. In \cite{wogrin2011generation,baringo2013strategic,jalal2015investment},  bilevel programs  are used to co-optimize generation expansion  decisions for different  types of power producers  in the upper level and wholesale market-clearing decisions in the lower level. Next, these programs are equivalently reformulated as a mathematical program with equilibrium constraints (MPEC) and converted into mixed-integer linear programs. To model a competition among multiple strategic market participants, models in \cite{kazempour2013generation,wogrin2012capacity,kwon2019resource} formulate an equilibrium problem with equilibrium constraints (EPEC), which consists of multiple MPECs. Each MPEC represents a strategic optimization problem of an individual market participant, while the EPEC seeks  a Generalized Nash Equilibrium (GNE) among all MPECs. In \cite{kazempour2013generation,wogrin2012capacity}, the GNE problem considers an energy market in the lower level, while \cite{kwon2019resource} extends the lower level to include  reserve and capacity markets. However, to our best knowledge, no work is devoted to analyzing regulatory competition in electricity markets by solving a stochastic EPEC.

\begin{figure}[t]
    \vspace{-0mm}
    \centering    
    \includegraphics[width=\columnwidth]{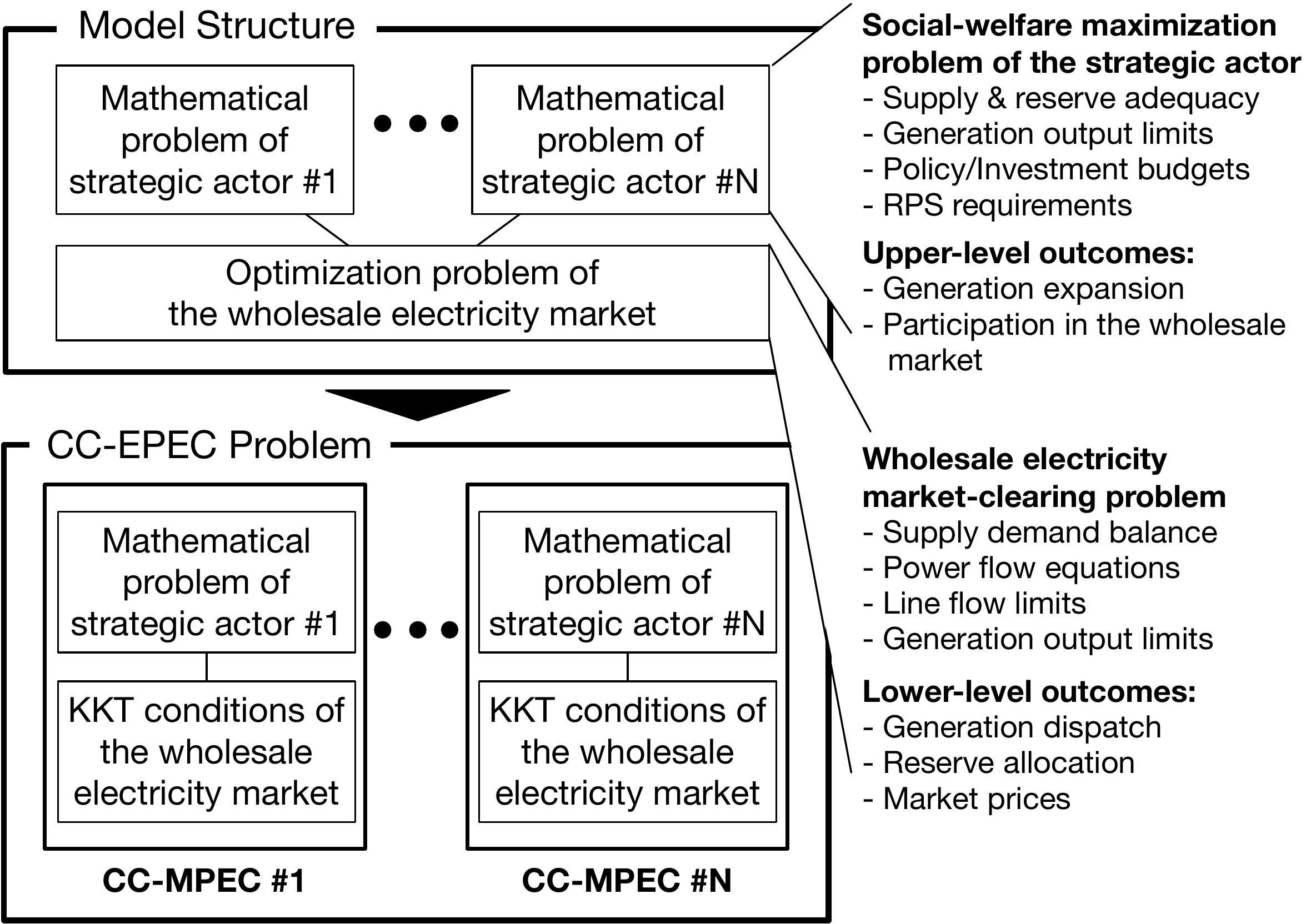} 
    \caption{\small Proposed models and reformulation structure.\vspace{-3mm}}
    \label{fig:model_structure}    
\end{figure}

This paper takes the \textit{combined perspective} of state regulators and in-state power companies to investigate effects of their \textit{coordinated} strategic behavior in a wholesale electricity market on achieving their RPS goals. This combined perspective is motivated because, in practice, state regulators and power companies jointly seek ways and means to achieve RPS goals \cite{utility2017}. Therefore, in our model, we assume that they coordinate generation expansion decisions, which in turn affects the power company's participation strategy  in the wholesale electricity market. To appropriately model these interactions in a multi-state regulatory environment, we first formulate a bilevel problem, later recast as an MPEC, for each strategic actor (i.e. the state regulator and in-state power companies) that aims to implement its RPS objective in the least-cost manner. Each bilevel problem considers a generation expansion problem in the upper level (including RES and fossil-fueled generation) and the wholesale electricity market in the lower level, which is common for all actors. To gain computational tractability, we model RES uncertainty via chance constraints and equivalently reformulate them as deterministic second-order conic constraints. In contrast to the traditional scenario-based stochastic models in \cite{wogrin2011generation,baringo2013strategic,jalal2015investment,kazempour2013generation,wogrin2012capacity,kwon2019resource}, which are computationally demanding and produce a solution that can be inefficient to individual scenarios considered, chance constrained stochastic programs yield robustified solutions toward  renewable uncertainties. Then, we formulate a chance constrained EPEC (CC-EPEC) to solve MPECs jointly using the lower-level KKT-optimality conditions as shown in Fig.~\ref{fig:model_structure}. The resulting problem is NP-hard due to the nonlinear and non-convex feasible region and, therefore, cannot be solved efficiently with off-the-shelf solvers, \cite{leyffer2010solving}. 

To overcome this computational complexity, we first linearize bilinear terms using the SOS1 variables and binary expansion approach \cite{pereira2005strategic}. Then, we customize the Progressive Hedging (PH) algorithm, \cite{rockafellar1991scenarios,watson2011progressive} to solve the proposed CC-EPEC. Unlike the conventional PH algorithm, which decomposes the original problem into scenario-dependent problems, our PH implementation decomposes the proposed CC-EPEC across each MPEC, i.e. we treat each strategic actor as a scenario, and solve them iteratively. Each decomposed MPEC is a mixed-integer second-order conic program (MISOCP) and can be tractably solved. 

Therefore, this paper makes two primary contributions. First, we propose a computationally tractable CC-EPEC formulation. Second, we develop a PH-inspired algorithm to efficiently solve this CC-EPEC, which otherwise cannot be solved with off-the-shelf solvers (our attempts failed due to the out-of-memory issues). These technical contributions make it possible to analyze the generation expansion decisions needed to support the RPS objectives in the ISO New England system with  sufficient reserve capacity provided by controllable generators under different retirement scenarios.

\section{Model}
The proposed model and its reformulation are structured in Fig.~\ref{fig:model_structure}.
Within the limited investment budget, the state regulator and power companies in each state devise a generation expansion plan to achieve a given RPS goal. This section first describes an uncertainty model and dispatch of controllable generation resources under uncertainty. Then, we formulate a CC-MPEC for each strategically acting actor and chance constraints are converted into the exact equivalent SOC formulation. Finally, given the CC-MPEC formulation, we model a regulatory competition problem as CC-EPEC.

\subsection{Uncertainty  model} 
Generation outputs of existing RES generation units $i\!\in\!{\cal I}^{\mathrm{R}}$ with given predicted output $\overline{g}_{ite}\!=\!\rho_{ite} G^{\mathrm{max}}_i$ and forecast error $\bm{\epsilon}_{ite}$ are modeled as a random variable:\vspace{-1mm}
\begin{subequations}\label{Eq:Random_model}\begin{align}    
    & \bm{g}_{ite} = \overline{g}_{ite} + \bm{\epsilon}_{ite} = \rho_{ite} G_i^{\mathrm{max}} + \bm{\epsilon}_{ite}, ~\forall i \in {\cal I}^{\mathrm{R}}, t \in {\cal T}, e\in{\cal E},\label{Eq:uncertainty_model}\vspace{-2mm}    
\end{align}
where $\rho_{ite}\in [0,1]$ is a forecast factor of RES generator $i$ and $G^{\mathrm{max}}_i$ is its installed capacity. The forecast error is assumed to be Gaussian, $\bm{\epsilon}_{ite} \sim \Norm\left({G^{\mathrm{max}}_i}\upsilon_{ite}, ({G^{\mathrm{max}}_i} \sigma_{ite})^2\right)$, with mean ${G^{\mathrm{max}}_i}\upsilon_{ite}$ and standard deviation $G^{\mathrm{max}}_i \sigma_i$, where $\upsilon_{ite}$ and $\sigma_{ite}$ are the normalized mean and standard deviation. 

Similarly, the generation output of candidate RES generation units $i\in\hat{\cal I}^{\mathrm{R}}$ is given in Eq.~\eqref{Eq:uncertainty_model_n} with investment decision variable $g^{\mathrm{max}}_{i}$ and forecast error $\bm{\epsilon}_{ite} \sim \mathrm{Norm}\left({g^{\mathrm{max}}_i} \upsilon_{ite}, ({g^{\mathrm{max}}_i} \sigma_{ite})^2\right)$:\vspace{-2mm}
\begin{align}    
    & \bm{g}_{ite} = \overline{g}_{ite} + \bm{\epsilon}_{ite}= \rho_{ite} g_i^{\mathrm{max}} + \bm{\epsilon}_{ite}, ~\forall i \in \hat{\cal I}^{\mathrm{R}}, t \in {\cal T}, e\in{\cal E}, \label{Eq:uncertainty_model_n}    
\end{align}
Since forecast errors result in a mismatch between the power produced and power consumed, existing and candidate controllable generation resources ($i\in{\cal I}^{\mathrm{C}}\cup\hat{\cal I}^{\mathrm{C}}$) must offset this imbalance. In practice, the real-time affine control is used \cite{jaleeli1992understanding} and modeled as:\vspace{-1mm}
\begin{align}
    & \bm{g}_{ite} = \overline{g}_{ite} \!-\! \alpha_{it} \!\!\!\!\!\!\!\!\!\!\sum_{j \in {\cal I}^{\mathrm R}_{s(i)}\cup\hat{\cal I}^{\mathrm R}_{s(i)}} \!\!\!\!\!\!\!\!\!\!\bm{\epsilon}_{jte},~~\forall i \in {\cal I}^{\mathrm C}\!\cup\!\hat{\cal I}^{\mathrm C}, t \in {\cal T}, e\in{\cal E}, \label{Eq:affine_control}
\end{align}
\end{subequations}
where $\sum_{i\in{\cal I}^{\mathrm C}_s\cup\hat{\cal I}^{\mathrm C}_s} \alpha_{it} \!=\! 1$. Synchronized generators continuously offset the power mismatch and control parameter $\alpha_{it}$ is usually set ahead of time, e.g. $\alpha_{it}=1/{\mathrm{card}({\cal{I}}^{\mathrm{C}}_s\cup \hat{\cal{I}}^{\mathrm{C}}_s})$, or can be optimized as in \cite{bienstock2014chance}.

\subsection{Optimization of a single strategic actor}
We formulate a bilevel program of the single strategic actor (state regulator and in-state power company) as:\hspace{-2mm}
\begin{subequations}\label{Eq:genco}
    \begin{align}        
        \begin{split}    
            & \max_{\Xi^{\mathrm{SA}}} O^{\mathrm{SA}}_s := \mathbb{E} \Big[ \sum_{e\in{\cal E}}\Big\{\omega_e\sum_{t\in{\cal T}} \Big(\!\!\sum_{n\in{\cal N}_s}\!\!\!\pi_{nt}^{\mathrm{D}}D_{nte} - \!\!\!\!\sum_{n\in{\cal N}_s}\!\!\!\lambda_{nte} p^{\downarrow}_{nte} \hspace{-10mm}
            \\& \hspace{20mm}- P^{\mathrm{ET}}_{s} \!\!\!\!\!\!\sum_{i\in{\cal I}_s^{\mathrm R}\cup\hat{\cal I}_s^{\mathrm R}} \!\!\!\!\!\!\bm{g}_{ite} - \!\!\sum_{i\in{\cal I}_s} C^{\mathrm{g}}_i \bm{g}_{ite} \Big)\Big\}\Big] 
            \\& \hspace{20mm}- P^{\mathrm{CT}}_s \!\!\sum_{i\in \hat{\cal I}^{\mathrm{R}}_s}  {g}^{\mathrm{max}}_{i}
            - \!\!\sum_{i\in \hat{\cal I}_s} C^{\mathrm{inv}}_i {g}^{\mathrm{max}}_{i}
              \label{Eq:genco_obj}
        \end{split}\\
        & \text{subject to:}\nonumber\\
        & \mathbb{P}[\bm{g}_{ite} \le {G}_{i}^{\mathrm{max}}]\ge 1-\eta,~\forall i \in {\cal I}^{\mathrm{C}}_s, t \in {\cal T}, e\in{\cal E}, \label{Eq:gen_UB}\\
        & \mathbb{P}[ \bm{g}_{ite} \le {g}_{i}^{\mathrm{max}}] \ge 1-\eta,~\forall i \in  \hat{\cal I}^{\mathrm{C}}_s, t \in {\cal T}, e\in{\cal E},\label{Eq:gen_UB_New}\\    
        & \mathbb{P}[ \bm{g}_{ite} \ge {G}_{i}^{\mathrm{min}} ]\ge 1-\eta,~\forall i \in  {\cal I}^{\mathrm{C}}_s, t \in {\cal T}, e\in{\cal E}, \label{Eq:gen_LB}\\
        & \mathbb{P}[ \bm{g}_{ite} \ge \Gamma_{i}{g}_{i}^{\mathrm{max}}] \ge 1-\eta,~\forall i \in  \hat{\cal I}^{\mathrm{C}}_s, t \in {\cal T}, e\in{\cal E},\label{Eq:gen_LB_New}\\    
        & {H}^{\mathrm{min}}_{i} \!\le\! \overline{g}_{ite}\!-\!\overline{g}_{i,t-1,e} \!\le\! {H}^{\mathrm{max}}_{i},~\forall i \in  {\cal I}^{\mathrm{C}}_s\!\cup\!\hat{\cal I}^{\mathrm{C}}_s, t \in {\cal T}, e\in{\cal E},\label{Eq:ramp}\\                
        & \sum_{i \in {\cal I}_n \cup {\hat{\cal{I}}_n}} \!\!\!\overline{g}_{ite}  + \!{p}_{nte}^{\downarrow} = \!{D}_{nte} ,\quad\forall t \in {\cal T},{n\in{\cal N}_s}, e\in{\cal E},\label{TSO_P_balance2}   \\                       
        & -\!{P}_n^{\downarrow,\mathrm{max}} \!\!\le {p}^{\downarrow}_{nte} \le {P}_n^{\downarrow,\mathrm{max}},~ \forall t \in {\cal T}\!, n\in{\cal N}_s, e\in{\cal E},\\
        & \sum_{t \in {\cal T}} \sum_{i \in {\cal I}_s^{\mathrm R}\cup \hat{\cal I}_s^{\mathrm R}} \!\!\overline{g}_{ite} \ge \kappa_s \sum_{t \in {\cal T}} \sum_{n\in{\cal N}_s} {D}_{nte},~\forall e\in{\cal E}, \label{Eq:StrReg_kappa}\\
        & \sum_{i\in\hat{\cal I}_s} C^{\mathrm{inv}}_i {g}^{\mathrm{max}}_{i}  \le  B^{\mathrm{C}}_s,\label{Eq:Cap_budget}\\    
        & \sum_{t \in {\cal T}} \Big(\! P^{\mathrm{ET}}_{s} \!\!\!\!\!\!\sum_{i\in{\cal I}_s^{\mathrm R}\cup\hat{\cal I}_s^{\mathrm R}} \!\!\!\!\overline{g}_{ite} \Big)+ P^{\mathrm{CT}}_{s}  \sum_{i\in\hat{\cal I}_s^{\mathrm R}} {g}^{\mathrm{max}}_{i}   \le  B^{\mathrm{P}}_s,~\forall e\in{\cal E},\label{Eq:Tariff_budget}\\
        \begin{split}            
            &\Big\{ \Xi^{\mathrm{WM}}_e\subseteq\arg\max \Big( O^{\mathrm{WM}}\!:=\!
            \sum_{t\in{\cal T}} \sum_{i \in {\cal I}\cup\hat{\cal I}} -C_i^{\mathrm{g}} {g}_{ite}\Big),\label{TSO_Obj}
        \end{split}\\          
        &(\xi_{lte}):\!\!\!\quad f_{lte} = \frac{1}{X_l} (\theta_{o(l),t,e}-\theta_{r(l),t,e}),~\forall {l \in {\cal L}},~t\in{\cal T},\label{TSO_DSPF}\\    
        \begin{split}
            &(\lambda_{nte}):\hspace{-1mm}\sum_{i \in {\cal I}_n\cup\hat{\cal I}_n } \hspace{-1.5mm}{g}_{ite}  + \hspace{-1mm}\sum_{l \vert r(l) = n } \hspace{-1.5mm} f_{lte} \!-\hspace{-1mm}\sum_{l \vert o(l) = n } \hspace{-1.5mm} f_{lte} = D_{nte}, 
            \\& \hspace{55mm}\forall n \in {\cal N}, t\in{\cal T},
        \end{split}\label{TSO_P_balance}\\
        &(\underline{\gamma}_{ite},\overline{\gamma}_{ite}):\!\quad 0 \le {g}_{ite} \le {g}^{\mathrm{o}}_{ite},~\forall i \in {\cal{I}}\cup\hat{\cal{I}},~t\in{\cal T}\!,\label{TSO_G_plim}\\
        & \text{where} ~ g_{ite}^{\mathrm{o}} \!=\!
        \begin{cases}            
            &\hspace{-3mm}\overline{g}_{ite},\hspace{20mm}\forall i\in{\cal I}_s\cup\hat{\cal I}_s, ~t\in{\cal T},\\
            &\hspace{-3mm}\rho_{ite}G^{\mathrm{max}}_{i} - R_{ite},~\forall i\in{\cal I}_{-s}\cup\hat{\cal I}_{-s}, ~t\in{\cal T},
        \end{cases}\nonumber
    \end{align}
    \begin{align}      
        &(\underline{\delta}_{lte},\overline{\delta}_{lte}):\!\quad -{F}^{\mathrm{max}}_l \le f_{lt} \le {F}^{\mathrm{max}}_l,~\forall {l \in {\cal L}},~t\in{\cal T}\!,\!\!\label{TSO_LineLim}\\
        & \Big\},\quad\forall e\in{\cal E},  \nonumber
        \end{align}
\end{subequations}
\noindent where $\Xi^{\mathrm{SA}} \!=\! \big\{ \overline{g}_{ite}, {g}^{\mathrm{max}}_i \!\geq\! 0;~p^{\downarrow}_{nte}\text{: free}\big\}$ and $\Xi^{\mathrm{WM}}_e \!=\! \big\{ {g}_{ite}, f_{lte}, \theta_{nte} \text{: free} \big\}$. 
The upper level in Eq.~\eqref{Eq:genco_obj}-\eqref{Eq:Tariff_budget} optimizes decisions of the single strategic actor, while the lower level in Eq.~\eqref{TSO_Obj}-\eqref{TSO_LineLim} clears the wholesale electricity market.

In the upper-level problem, objective function \eqref{Eq:genco_obj} maximizes the expected social-welfare of state $s$ over the operating horizon $t\in{\cal T}$, which includes the revenue collected by selling electricity in-state minus the operating and investment costs. Note that investment cost $C^{\mathrm{inv}}_i$ is pro-rated on a daily basis using the net present value approach as in \cite{6936940}.
Eq.~\eqref{Eq:gen_UB}-\eqref{Eq:gen_LB_New} enforce the power output limits of existing (${\cal I}^{\mathrm{C}}_s$) and candidate controllable generators ($\hat{\cal I}^{\mathrm{C}}_s$) of the strategic actor in state $s$, while the uncertainty imposed by renewable resources is modeled via chance constraints, and random variable $\bm{g}_{ite}$ is defined as in Eq.~\eqref{Eq:Random_model}.
Ramping constraints of controllable generators are given in \eqref{Eq:ramp}, and Eq.~\eqref{Eq:StrReg_kappa} imposes the RPS requirement in state $s$. 
The capital cost budget is enforced in Eq.~\eqref{Eq:Cap_budget} and the budget limits on the renewable policy, including energy feed-in-tariff $P^{\mathrm{ET}}_s$ and capacity tariff $P^{\mathrm{CT}}_s$, are given in Eq.~\eqref{Eq:Tariff_budget}.
In the lower-level problem, objective function \eqref{TSO_Obj} maximizes the social welfare across all states $s\in{\cal S}$.
Eq.~\eqref{TSO_DSPF} models DC power flows and Eq.~\eqref{TSO_P_balance} enforces the power balance.
Eq.~\eqref{TSO_G_plim}-\eqref{TSO_LineLim} limit  generation outputs and line capacities. Dual variables of Eq.~\eqref{TSO_DSPF}-\eqref{TSO_LineLim} are defined in parentheses before each constraint.

\subsection{Conic reformulation of the chance constraints}
Using \cite{bienstock2014chance,lubin2019chance}, the chance constraints in Eq.~\eqref{Eq:gen_UB}-\eqref{Eq:gen_LB_New} can be equivalently reformulated into deterministic constraints \eqref{Eq:gen_UB_Conic}-\eqref{Eq:gen_LB_New_Conic}, where $\Phi^{-1}(\cdot)$ is an inverse cumulative distribution function and $\Stdev[\cdot]$ is a standard deviation operator:
\begin{subequations}\label{Eq:ConicReform}
    \begin{align}          
        \begin{split}
            & \overline{g}_{ite} 
            -\alpha_{it} \Big( \hspace{-0.0mm} \sum_{j\in{\cal I}^{\mathrm{R}}_{s}} \hspace{-1.0mm} G^{\mathrm{max}}_j \upsilon_{jte} 
            \!+ \hspace{-1.0mm} \sum_{j\in\hat{\cal I}^{\mathrm{R}}_{s}} \hspace{-1.0mm} g^{\mathrm{max}}_j \upsilon_{jte}\Big)
            \\& \hspace{5.5mm}+ \Phi^{-1}(1\!-\!\eta){\Stdev[\bm{g}_{it}]} 
            \le {G}_{i}^{\mathrm{max}},~\forall i \in  {\cal I}^{\mathrm{C}}_s, t \in {\cal T}, 
        \end{split}\label{Eq:gen_UB_Conic}\\
        \begin{split}
            & \overline{g}_{ite} 
            -\alpha_{it} \Big( \hspace{-0.0mm} \sum_{j\in{\cal I}^{\mathrm{R}}_{s}} \hspace{-1.0mm} G^{\mathrm{max}}_j \upsilon_{jte} 
            \!+ \hspace{-1.0mm} \sum_{j\in\hat{\cal I}^{\mathrm{R}}_{s}} \hspace{-1.0mm} g^{\mathrm{max}}_j \upsilon_{jte}\Big)
            \\& \hspace{5.5mm}+ \Phi^{-1}(1\!-\!\eta){\Stdev[\bm{g}_{it}]} 
            \le {g}_{i}^{\mathrm{max}},~\forall i \in  \hat{\cal I}^{\mathrm{C}}_s, t \in {\cal T}, 
        \end{split}\label{Eq:gen_UB_New_Conic}\\    
        \begin{split}
            & \overline{g}_{ite} 
            -\alpha_{it} \Big( \hspace{-0.0mm} \sum_{j\in{\cal I}^{\mathrm{R}}_{s}} \hspace{-1.0mm} G^{\mathrm{max}}_j \upsilon_{jte} 
            \!+ \hspace{-1.0mm} \sum_{j\in\hat{\cal I}^{\mathrm{R}}_{s}} \hspace{-1.0mm} g^{\mathrm{max}}_j \upsilon_{jte}\Big)
            \\& \hspace{5.5mm}- \Phi^{-1}(1\!-\!\eta){\Stdev[\bm{g}_{it}]} 
            \ge {G}_{i}^{\mathrm{min}},~\forall i \in  {\cal I}^{\mathrm{C}}_s, t \in {\cal T}, 
        \end{split}\label{Eq:gen_LB_Conic}\\
        \begin{split}
            & \overline{g}_{ite} 
            -\alpha_{it} \Big( \hspace{-0.0mm} \sum_{j\in{\cal I}^{\mathrm{R}}_{s}} \hspace{-1.0mm} G^{\mathrm{max}}_j \upsilon_{jte} 
            \!+ \hspace{-1.0mm} \sum_{j\in\hat{\cal I}^{\mathrm{R}}_{s}} \hspace{-1.0mm} g^{\mathrm{max}}_j \upsilon_{jte}\Big)
            \\& \hspace{5.5mm}- \Phi^{-1}(1\!-\!\eta){\Stdev[\bm{g}_{it}]} 
            \ge \Gamma_i {g}_{i}^{\mathrm{max}},~\forall i \in  \hat{\cal I}^{\mathrm{C}}_s, t \in {\cal T}, \hspace{-10mm}
        \end{split}\label{Eq:gen_LB_New_Conic}\\
        & \Stdev[\bm{g}_{it}] \!=\! \alpha_{it}\!\sqrt{\sum_{j\in{\cal{I}}_{s(i)}^{\mathrm{R}}}({{G}^{\mathrm{max}}_{j}} \sigma_{jte})^2 \!+\!\! \sum_{j\in\hat{\cal{I}}_{s(i)}^{\mathrm{R}}}({{g}^{\mathrm{max}}_{j}} \sigma_{jte})^2 }. \label{Eq:stdev}
    \end{align}
\end{subequations}

Since participation factor $\alpha_{it}$ can be modeled as a parameter \cite{jaleeli1992understanding}, Eq.~\eqref{Eq:ConicReform} can be represented as second-order conic constraints. We define auxiliary vector variable $x_{ste}\in{\mathbb{R}}^{u+v}$ in \eqref{Eq:x_def}, where $u\!=\! \mathrm{card}({\cal{I}}^{\mathrm{R}}_s)$, and $v\!=\! \mathrm{card}(\hat{\cal{I}}^{\mathrm{R}}_s)$, so that $\Stdev[\bm{g}_{it}]= \alpha_{it} \|x_{s(i),t,e} \|_2$. The remaining terms in Eq.~\eqref{Eq:ConicReform}, except for $\| x_{s(i),t,e} \|_2$, are aggregated in variables $\underline{y}_{ite}, \overline{y}_{ite} \in \mathbb{R}$ as shown in Eq.~\eqref{Eq:y_def_LB}-\eqref{Eq:y_def_UB}: 
\begin{subequations}\label{Eq:xydef}
    \begin{align} 
        \begin{split}       
        &\hspace{-4mm} x_{s(i),t,e} \!\!:=\! 
            \big[ {{G}^{\mathrm{max}}_{1}} \sigma_{1te},\cdots,{{G}^{\mathrm{max}}_{u}} \sigma_{ute},~{{g}^{\mathrm{max}}_{1}} \sigma_{u+1,t,e},
            \\&\qquad\qquad \cdots,{{g}^{\mathrm{max}}_{v}} \sigma_{u+v,t,e}  \!\big]^{\!\top\!},~\forall i \in {\cal I}^{\mathrm C}\cup\hat{\cal I}^{\mathrm C}
        \end{split}\label{Eq:x_def}\\
        & \hspace{-4mm}\underline{y}_{ite} \hspace{-1.1mm}:=\hspace{-1.3mm}
            \begin{cases}             
                \hspace{-0.4mm}\frac{
                \{
                \overline{g}_{ite} 
                -\alpha_{it} ( \hspace{-0.0mm} \sum_{j\in{\cal I}^{\mathrm{R}}_{s(i)}} \hspace{-1.0mm} G^{\mathrm{max}}_j \upsilon_{jte} 
                \!+ \hspace{-0.0mm} \sum_{j\in\hat{\cal I}^{\mathrm{R}}_{s(i)}} \hspace{-1.0mm} g^{\mathrm{max}}_j \upsilon_{jte})\}
                -G^{\mathrm{min}}_i}{\Phi^{-1}(1-\eta)\alpha_{it}} ,\!\!
                \\
                \hspace{32.5mm}\forall i \in  {\cal I}^{\mathrm{C}}_s,~t \in {\cal T},~e\in{\cal E},
                \\                
                \hspace{-0.4mm}\frac{
                \!\!\{
                \overline{g}_{ite} 
                -\alpha_{it} ( \hspace{-0.0mm} \sum_{j\in{\cal I}^{\mathrm{R}}_{s(i)}} \hspace{-1.0mm} \!\!G^{\mathrm{max}}_j \upsilon_{jte} 
                \!+ \hspace{-0.0mm} \sum_{j\in\hat{\cal I}^{\mathrm{R}}_{s(i)}} \hspace{-1.0mm} g^{\mathrm{max}}_j \upsilon_{jte})\}
                -\Gamma_i g^{\mathrm{max}}_i}{\Phi^{-1}(1-\eta)\alpha_{it}} ,
                \\\hspace{32.5mm}\forall i \in  \hat{\cal I}^{\mathrm{C}}_s, ~t \in {\cal T},~e\in{\cal E},
            \end{cases}\hspace{-6mm} \label{Eq:y_def_LB}\\    
        & \hspace{-4mm}\overline{y}_{ite} \hspace{-1.1mm}:=\hspace{-1.3mm}
            \begin{cases}                 
                \hspace{-0.4mm}\frac{
                G^{\mathrm{max}}_i - 
                \{
                \overline{g}_{ite} 
                -\alpha_{it} ( \hspace{-0.0mm} \sum_{j\in{\cal I}^{\mathrm{R}}_{s(i)}} \hspace{-1.0mm} G^{\mathrm{max}}_j \upsilon_{jte} 
                \!+ \hspace{-0.0mm} \sum_{j\in\hat{\cal I}^{\mathrm{R}}_{s(i)}} \hspace{-1.0mm} g^{\mathrm{max}}_j \upsilon_{jte})\}
                }
                {\Phi^{-1}(1-\eta)\alpha_{it}} ,\hspace{-10mm}
                \\\hspace{32.5mm}\forall i \in  {\cal I}^{\mathrm{C}}_s,~t \in {\cal T},~e\in{\cal E},\hspace{-10mm}
                \\
                \hspace{-0.4mm}\frac{
                g^{\mathrm{max}}_i - 
                \{
                \overline{g}_{ite} 
                -\alpha_{it} ( \hspace{-0.0mm} \sum_{j\in{\cal I}^{\mathrm{R}}_{s(i)}} \hspace{-1.0mm} G^{\mathrm{max}}_j \upsilon_{jte} 
                \!+ \hspace{-0.0mm} \sum_{j\in\hat{\cal I}^{\mathrm{R}}_{s(i)}} \hspace{-1.0mm} g^{\mathrm{max}}_j \upsilon_{jte})\}
                }
                {\Phi^{-1}(1-\eta)\alpha_{it}} ,\hspace{-10mm}
                \\\hspace{32.5mm}\forall i \in  \hat{\cal I}^{\mathrm{C}}_s,~t \in {\cal T},~e\in{\cal E}.\hspace{-10mm}
            \end{cases}\hspace{-10mm} \label{Eq:y_def_UB}
    \end{align}
\end{subequations}

Finally, using Eq.~\eqref{Eq:ConicReform}-\eqref{Eq:xydef}, we can replace chance constraints in Eq.~\eqref{Eq:gen_LB}-\eqref{Eq:gen_UB_New} by deterministic SOC constraints as follows: 
\begin{subequations}\label{Eq:SOC_refined}
    \begin{align}              
        &  \begin{bmatrix}\underline{y}_{ite}\\x_{s(i),t,e}\end{bmatrix}\! \in {\cal K} ^{u+v+1},\quad\forall i \in {\cal I}_s^{\mathrm C}\cup\hat{\cal I}_s^{\mathrm C},t \in {\cal T},e \in {\cal E},\!\!\label{Eq:SOC_primal_feasibility1}\\
        & \begin{bmatrix}\overline{y}_{ite}\\x_{s(i),t,e}\end{bmatrix}\! \in {\cal K} ^{u+v+1},\quad\forall i \in {\cal I}_s^{\mathrm C}\cup\hat{\cal I}_s^{\mathrm C},t \in {\cal T},e \in {\cal E},\!\!\label{Eq:SOC_primal_feasibility2} 
    \end{align}
 where ${\cal K}^{n+1}$ is the following second-order cone:
 \begin{align}
     &{\cal K}^{n+1}\! :=\! \big\{ \!\begin{bmatrix}y;~x\end{bmatrix} \!\!~\vert~\! y \!\ge\! \sqrt{x_1^2 \! +\! \cdots\! +\! x_n^2, }~y\!\in\!{\mathbb R},  x\!\in\!{\mathbb R}^n \!\big\}.
 \end{align}\end{subequations}

\subsection{CC-MPEC reformulation}\label{subsec:MPEC_reform}
Since the lower-level problem in \eqref{TSO_Obj}-\eqref{TSO_LineLim} is linear, Eq.~\eqref{Eq:genco} can be equivalently converted into a single-level CC-MPEC problem by using the lower-level KKT optimality conditions:
\begin{subequations}\label{MPEC}\begin{align}    
    \begin{split}    
        & \max_{\Xi^{\mathrm{SA}}} O^{\mathrm{SA}}_s := \mathbb{E} \Big[ \sum_{e\in{\cal E}}\Big\{\omega_e\sum_{t\in{\cal T}} \Big(\!\!\sum_{n\in{\cal N}_s}\!\!\!\pi_{nt}^{\mathrm{D}}D_{nte} - \!\!\!\!\sum_{n\in{\cal N}_s}\!\!\!\lambda_{nte} p^{\downarrow}_{nte} \hspace{-10mm}
        \\& \hspace{20mm}- P^{\mathrm{ET}}_{s} \!\!\!\!\!\!\sum_{i\in{\cal I}_s^{\mathrm R}\cup\hat{\cal I}_s^{\mathrm R}} \!\!\!\!\!\!\bm{g}_{ite} - \!\!\sum_{i\in{\cal I}_s} C^{\mathrm{g}}_i \bm{g}_{ite} \Big)\Big\}\Big] 
        \\& \hspace{20mm}- P^{\mathrm{CT}}_s \!\!\sum_{i\in \hat{\cal I}^{\mathrm{R}}_s}  {g}^{\mathrm{max}}_{i}
        - \!\!\sum_{i\in \hat{\cal I}_s} C^{\mathrm{inv}}_i {g}^{\mathrm{max}}_{i}
          \label{Eq:genco_obj2}
    \end{split}\\
    & \text{subject to:}\nonumber\\
    & \text{Upper-level Constraints}: ~\hspace{13.5mm} \text{Eq.}~\eqref{TSO_P_balance2}\textendash\eqref{Eq:Tariff_budget},\eqref{Eq:SOC_refined}\label{Eq:GenconCon}\\
    & \text{Lower-level Equality Constraints}:~\text{Eq.}~\eqref{TSO_DSPF}\textendash\eqref{TSO_P_balance}\label{Eq:TSO_EQcons}\\
    & \text{Lower-level KKT conditions:}\nonumber\\    
    \begin{split}
        & -  C^{\mathrm{g}}_i+\lambda_{n(i),t,e} + \underline{\gamma}_{ite} - \overline{\gamma}_{ite} = 0,
        \\& \hspace{43mm}\forall {i \in {\cal I}\cup\hat{\cal I}},t\in{\cal T},e\in{\cal E},\hspace{-20mm}
    \end{split}\label{Eq:KKT_gp}\\
    \begin{split}
        & \xi_{lte}+\lambda_{r(l),t,e}-\lambda_{o(l),t,e}+\underline{\delta}_{lte}-\overline{\delta}_{lte}  = 0,
        \\&\hspace{43mm}\forall l \in {\cal{L}},~t\in{\cal T},~e\in{\cal E},
    \end{split}\label{Eq:KKT_fp}\\
    & -\hspace{-3mm}\sum_{l \vert o(l) = n}\hspace{-2mm} \frac{\xi_{lte}}{X_l}+\!\!\!\!\sum_{l \vert r(l) = n}\hspace{-2mm} \frac{\xi_{lte}}{X_l} = 0, \quad\forall n \in {\cal{N}},~t\in{\cal T},~e\in{\cal E},\label{Eq:KKT_theta}
\end{align}
\begin{align}
    & 0 \le g_{ite} \perp \underline{\gamma}_{ite} \ge 0,\quad\forall i \in {\cal I}\cup\hat{\cal I},~t\in{\cal T},~e\in{\cal E},\label{Eq:KKT_gammaLB}\\
    & 0 \le {g}^{\mathrm o}_{ite} \!-\! g_{ite} \!\perp \!\overline{\gamma}_{ite}\! \ge 0,\quad\forall i \in {\cal I}\cup\hat{\cal I},~t\in{\cal T},~e\in{\cal E},\label{Eq:KKT_gammaUB}\\
    & 0 \le f_{lte}-{F}^{\mathrm{min}}_l \perp \underline{\delta}_{lte} \ge 0,\quad\forall l \in {\cal L},~t\in{\cal T},~e\in{\cal E},\label{Eq:KKT_deltaLB}\\
    & 0 \le {F}^{\mathrm{max}}_l-f_{lte} \perp \overline{\delta}_{lte} \ge 0,\quad\forall l \in {\cal L},~t\in{\cal T},~e\in{\cal E},\label{Eq:KKT_deltaUB}
\end{align}

\end{subequations}
\noindent where Eq.~\eqref{Eq:GenconCon} lists constraints of the strategic actor, Eq.~\eqref{Eq:TSO_EQcons} restates the lower-level equality constraints and Eq.~\eqref{Eq:KKT_gp}-\eqref{Eq:KKT_deltaUB} describe the stationary conditions and complementary slackness of lower-level inequality constraints. Note that $\perp$ in Eq.~\eqref{Eq:KKT_gammaLB}-\eqref{Eq:KKT_deltaUB} denotes orthogonality, i.e. $x\perp y \Leftrightarrow x^{\top}y=0$.

\vspace{-1mm}
\subsection{CC-EPEC formulation}
Finally, we formulate the CC-EPEC as: 
\vspace{-1.5mm}\begin{align}
    & \big\{[\text{CC-MPEC}^{\mathrm{SA}}_s~:=~\text{Eq.}~\eqref{MPEC}],~\forall {s\in{\cal S}}\big\},\vspace{-2mm}\label{Eq:CC-EPEC}
\end{align}
where $\text{CC-MPEC}^{\mathrm{SA}}_s$ represents the strategic actor (state regulator and power company) in state $s$ modeled in Section \ref{subsec:MPEC_reform}.

\vspace{2mm}
\section{Solution Technique}
The proposed CC-EPEC in \eqref{Eq:CC-EPEC} is computationally challenging due to the nonlinear and non-convex feasible region and two-stage (investment and operation) structure. Therefore, we first linearize bilinear terms to recast the CC-MPEC in \eqref{MPEC} as mixed-integer SOCP and then apply the PH algorithm.

\subsection{Linearization}\label{subsec:Lin}
The KKT conditions in the lower-level problem of Eq.~\eqref{Eq:CC-EPEC} include bilinear terms $g_{ite} \underline{\gamma}_{ite}, g_{ite}\overline{\gamma}_{ite}, f_{lte}\underline{\delta}_{lte}$ and $f_{lte}\overline{\delta}_{lte}$ in \eqref{Eq:KKT_gammaLB}\textendash\eqref{Eq:KKT_deltaUB}, which arise from complementarity slackness conditions. 
These nonlinearities are typically dealt with using the Fortuny-Amat \& McCarl transformation \cite{fortuny1981representation}, which replaces the bilinear terms with mixed-integer linear constraints parameterized by exogenous parameter (so-called big \textit{M}).
However, finding a value of big \textit{M} that ensures a stable and tractable computational performance is itself a NP-hard problem \cite{pineda2019solving}. In practice, the big \textit{M} method increases complexity of the underlying problem and leads to numerical issues even when state-of-the-art solvers are used \cite{siddiqui2013sos1,pineda2019solving,gurobi2018manual}. These challenges can be overcome by means of using SOS1 variables (a set of variables, in which at most one variable within the set is allowed to attain a non-zero value), which are numerically robust and supported  by off-the-shelf MIP solvers (e.g. Gurobi, CPLEX) \cite{cplex2018manual,gurobi2018manual}. For instance, $0\le g_{ite} \perp \underline{\gamma}_{ite} \ge 0$ in \eqref{Eq:KKT_gammaLB} can be recast as:
\begin{align}\label{Eq:SOS1}
    & \{g_{ite}, ~\underline{\gamma}_{ite}\} \in \text{SOS1},\quad    g_{ite}, \underline{\gamma}_{ite} \ge 0.
\end{align}
In fact, the advantage of \eqref{Eq:SOS1} relative to the Fortuny-Amat transformation is that MIP solvers (e.g. Gurobi)  reformulate SOS1 constraints as disjunctive constraints and validate their exactness to avoid numerical errors \cite{gurobi2018manual}. 

Additionally, the objective function in \eqref{Eq:genco_obj2} also includes bilinear term $\lambda_{n(i),t,e} g_{ite}$, which cannot be modeled via SOS1 variables because both $\lambda_{n(i),t,e}$ and $g_{ite}$ can  simultaneously attain non-zero values. Instead, we linearize  $\lambda_{n(i),t,e} g_{ite}$ using the binary expansion method, \cite{pereira2005strategic}, which  introduces auxiliary binary variables similar to the Fortuny-Amat \& McCarl transformation, but does not require orthogonality in bilinear terms. First, continuous variable $\lambda_{nte}$ is discretized into $2^K$ levels, where $K$ is a user-defined parameter, as:
\begin{subequations}\label{Eq:BEA}\begin{align}    
    & \lambda_{nte} = \underline{\lambda}_{n} + \Delta \lambda \sum_{k=1}^{K} 2^{k-1} z_{ntke}, \quad z_{ntke}\in \{0,1\}, \label{BEA:step1}
\end{align}\vspace{-1mm}
Then, bilinear term $\lambda_{n(i),t,e} g_{ite}$ can be written as:
\vspace{-2mm}\begin{align}
    & \lambda_{n(i),t,e}g_{ite} =\underline{\lambda}_{n(i)} g_{ite} + \Delta\lambda \sum_{k=1}^{K} 2^{k-1} s_{itke},\label{BEA:step2}\\\vspace{-1mm}
    &\hspace{-8mm} \text{where:}\nonumber\\\vspace{-1mm}
    & 0 \le s_{itke} \le g^{\mathrm{max}}_{i} z_{n(i),tke},\label{BEA:step4}\\
    & g_{ite}\!-g^{\mathrm{max}}_{i}(1\!-\!z_{n(i),t,k,e}) \le s_{n(i),t,k,e} \le g_{ite}.\label{BEA:step5}
\end{align}
\end{subequations}
There is a natural trade-off between the choice of discretization parameter $K$ and the accuracy of the binary expansion approach. This choice affects performance of the solver because it changes the number of discrete decisions and, hence, parameter $K$ should be calibrated carefully.

\subsection{PH Algorithm}\label{subsec:PH}
\begin{figure}[!b]
\vspace{-0mm}
    \centering    
    \includegraphics[width=0.94\columnwidth]{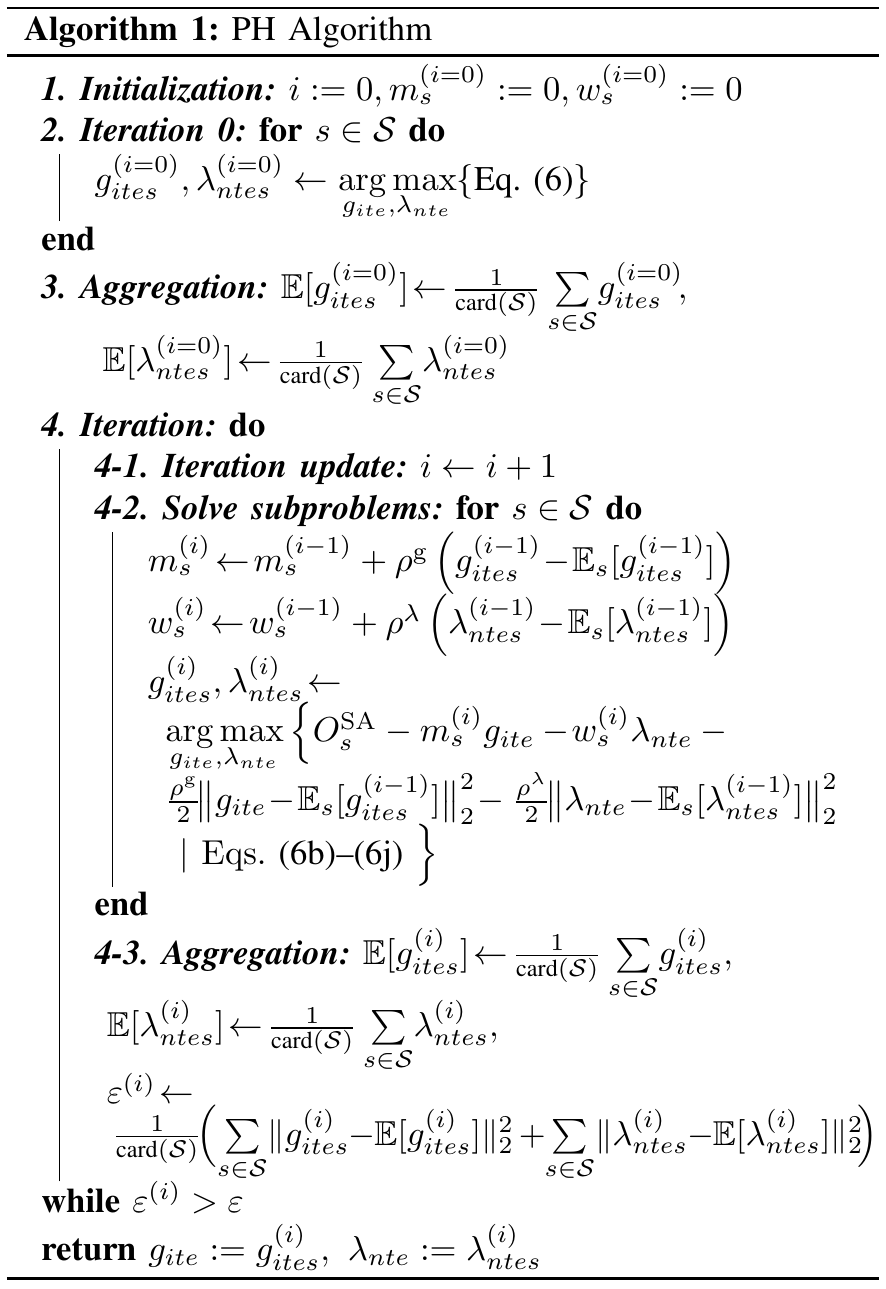}    
    \vspace{-0mm} 
    \label{alg:PH_algorithm} 
\end{figure}

Structurally, the  CC-EPEC in Eq.~\eqref{Eq:CC-EPEC} is a collection of CC-MPECs that share the same lower-level problem, which can be cast as a multi-leader-common-follower (MLCF) game, \cite{leyffer2010solving}, where leaders are strategic actors and the follower is the wholesale market. This MLCF representation motivates the decomposition of  the  CC-EPEC in Eq.~\eqref{Eq:CC-EPEC} for each leader and treat each decomposed problem as an individual ``scenario''. This leader-based decomposition, resembling a scenario-based two-stage stochastic optimization problem, makes it possible to customize the PH algorithm \cite{rockafellar1991scenarios}. Thus, lower-level decision variables $g_{its}$ and $\lambda_{nts}$ of each MPEC are treated as hedging variables. At each iteration, all decomposed MPECs are solved independently and the deviation of hedging variables from the consensus is penalized. The process iterates until the consensus is reached among hedging variables (the difference is sufficiently small), thus achieving an equilibrium among MPECs, which in solves the CC-EPEC in Eq.~\eqref{Eq:CC-EPEC}.

This  PH implementation is detailed in Algorithm 1. Step 1 initializes iteration counter $i$ and PH multipliers $m_s, \omega_s$. Step 2 executes the initial iteration and obtains the solution of each $\text{MPEC}_s^{\mathrm{SA}}$ in Eq.~\eqref{MPEC}. Step 3 computes the average of hedging variables. Next, Step 4-2 at each iteration  augments and solves  $\text{MPEC}_s^{\mathrm{SA}}$ in Eq.~\eqref{MPEC}, where the difference between a given and average hedging variable is penalized by PH multipliers $m_s^{(i)}$ and $w_s^{(i)}$, directing the solution  toward the consensus. Step 4-3 updates the value of hedging variables and computes achieved   termination tolerance $\epsilon^{(i)}$. Step 4 iterates until  a desired termination tolerance is achieved.

\section{Case Study}

\begin{table}[b]
\vspace{0mm}
    \centering
    \captionsetup{justification=centering, labelsep=period, font=footnotesize, textfont=sc}
    \caption{Existing generation capacity by state and type [MW]}        
    \begin{tabular}{P{16mm}|P{7mm} P{7mm} P{7mm} P{7mm} P{7mm} P{7mm}}    
        \hline
        \hline
        \multirow{2}{*}{}&ME&NH&VT&MA&CT&RI\\
        \hline
        Wind&221.2&140.5&39.0&681.7&132.5&85\\
        Solar&41.4&83.84&306.3&1871.26&464.34&116.66\\
        Nuclear&0&1244&620.2&684.7&2116&0\\
        Coal&311.8&95.4&0&144.4&744.4&1099.5\\
        Oil&1146.9&400.2&0&1111.7&2212.8&435\\
        Natural Gas&3862.7&508&0&2249.6&621.4&3491.6\\
        \hline
        \hline              
    \end{tabular}
    \label{table:gencap}              
    \vspace{-0mm}
\end{table}

\begin{table}[b]
\vspace{-0mm}
    \centering
    \captionsetup{justification=centering, labelsep=period, font=footnotesize, textfont=sc}
    \caption{Transmission lines specifications of the ISO New England system \cite{krishnamurthy20168}}        
    \begin{tabular}{P{5mm}|P{9mm} P{9mm} P{9mm} P{9mm} P{9mm} P{9mm}}        
        \hline
        \hline
        \multirow{2}{*}{Line}& \multirow{2}{*}{From}& \multirow{2}{*}{To}& Distance [miles]& Resistance [$\Omega$/m]& Reactance [$\Omega$/m]&Capacity [MW]\\
        \hline
        1&ME&NH&115&19.09&54.05&1200\\
        2&VT&NH&100&16.6&47&1200\\
        3&VT&WCMA&150&24.9&70.5&1200\\
        4&WCMA&NH&86&14.28&40.42&1200\\
        5&NEMA&WCMA&80&13.28&37.6&1200\\
        6&NEMA&NH&63&10.46&29.61&1200\\
        7&NEMA&SEMA&30&4.98&14.1&1200\\
        8&WCMA&CT&30&4.98&14.1&1200\\
        9&WCMA&RI&65&10.79&30.55&1200\\
        10&NEMA&RI&40&6.64&18.8&1200\\
        11&CT&RI&64&10.62&30.08&1200\\
        12&SEMA&RI&20&3.32&9.4&1200\\
        \hline
        \hline              
    \end{tabular}
    \label{table:line}              
    \vspace{-0mm}
\end{table}
We use the 8-zone ISO New England system, \cite{krishnamurthy20168}, covering 6 states, shown in Fig.~\ref{fig:platform}. 
The system includes 76 controllable and 16 renewable generation resources, and candidate generator set $\hat{I}_n$ contains natural gas ($C_i^{\mathrm{inv}}$=\$895/kW, $C_i^{\mathrm{g}}$=\$20/MWh), wind ($C_i^{\mathrm{inv}}$=\$1,630/kW, $C_i^{\mathrm{g}}$=\$1.1/MWh) and solar ($C_i^{\mathrm{inv}}$=\$2,434/kW, $C_i^{\mathrm{g}}$=\$0.4/MWh) options in each zone $n\in{\cal N}$. 
The installed capacity of existing resources in the system is itemized  by state and type in Table~\ref{table:gencap} and the transmission network data is provided in Table~\ref{table:line}.
The average system peak demand over all scenarios is $\sum_{e\in{\cal E}} \omega_s \sum_{n\in{\cal N}} D_{nte}=10,243~\mathrm{MW}$.
Capital and operating costs are calculated based on the data from the EIA 2019 annual report \cite{EIA_cost} and the wind and solar installed capacity and hourly forecast data is obtained from the ISO New England online library \cite{van2018state,isone2019final}.
In this case study,  we implement the RPS targets as given in Fig.~\ref{fig:platform} over the same planning horizon. 
All controllable generators are assumed to be committed and we set the normalized standard deviation of forecast errors ($\sigma_{ite}$) as 0.1 and 0.2 for wind and solar resources, while security tolerance is set to $\eta =0.03$. 
The net present value of investment costs assumes the cost recovery period of 10 years with 5\% annual discount rate. The PH penalty factors are set as $\rho^{g}\!=\!\rho^{\lambda}\!=\!0.7$ and dual variable $\lambda_{nte}$ is discretized with $K\!=\!10$ in Eq.~\eqref{BEA:step1}.  
The termination tolerance of the PH algorithm is $\epsilon=0.03$. 
Energy- and capacity-specific incentives are set to $P^{\mathrm{ET}}_s\!=\!\$3/\text{MWh}$ and $P^{\mathrm{CT}}_s\!=\!\$300/\text{kW}$ \cite{EIA_cost}. We consider different retirement scenarios: a) Basecase (no retirement), b) Coal retirement and c) Coal \& Nuclear retirement. The centralized generation planning model in \cite{munoz2014engineering}, detailed in Appendix, is used for benchmarking. All simulations are implemented using Julia v1.2.0 and JuMP v0.20 \cite{DunningHuchetteLubin2017} and solved by Gurobi v8.1 on an Intel Xeon 2.6 GHz processor with 10 cores and 250GB of memory. The code source and input data are available in \cite{julia_code_gist}.

\begin{figure*}[t!]
\centering
\makebox[0pt][c]{%
\hspace{-0mm}\begin{minipage}[b]{0.49\linewidth}
\centering
\includegraphics[scale=0.118]{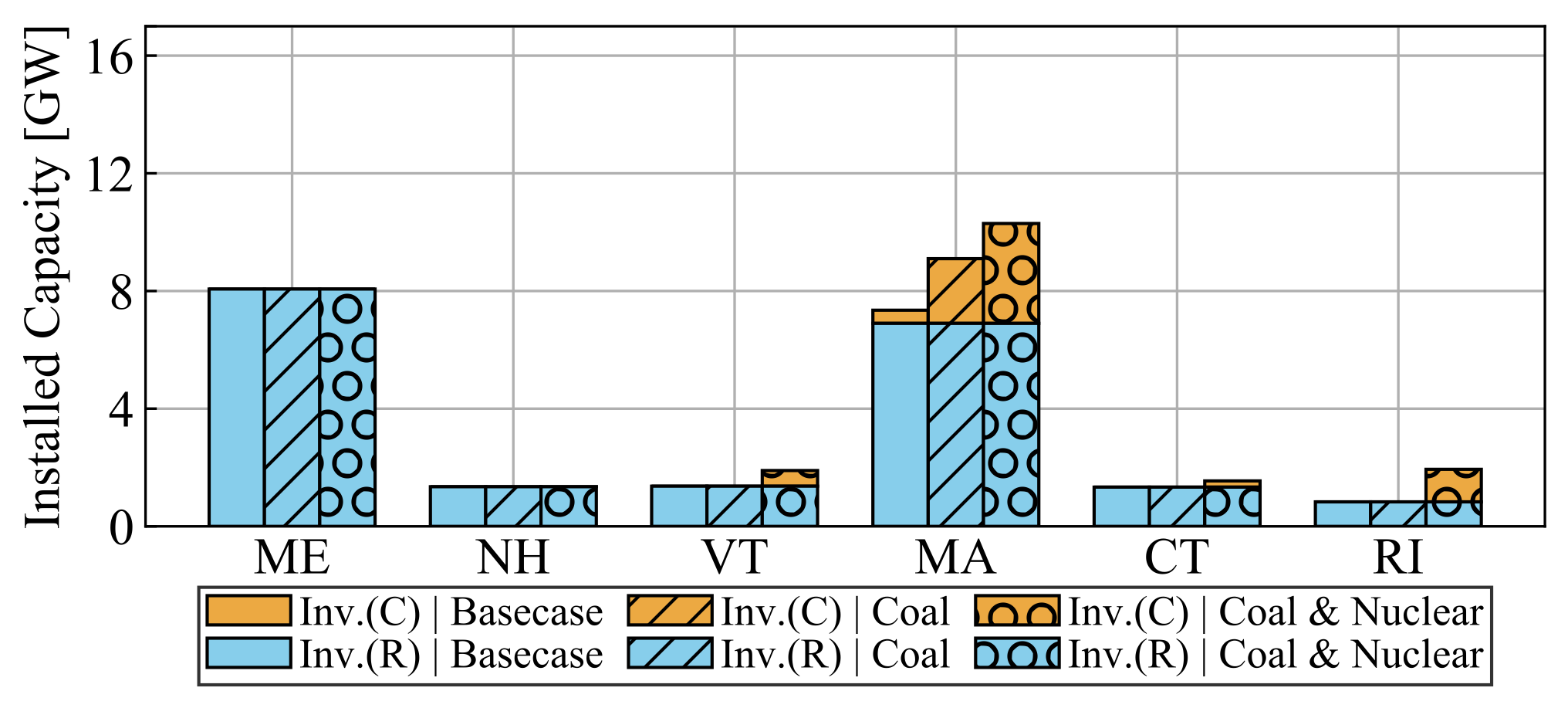}
    \vspace{-3mm}
    \caption{\small CC-MPEC: Generation expansion decisions for controllable (C) and renewable (R) resources under different retirement scenarios. \vspace{-1.5mm}}
    \label{fig:MPEC_gpmaxF}    
\includegraphics[scale=0.118]{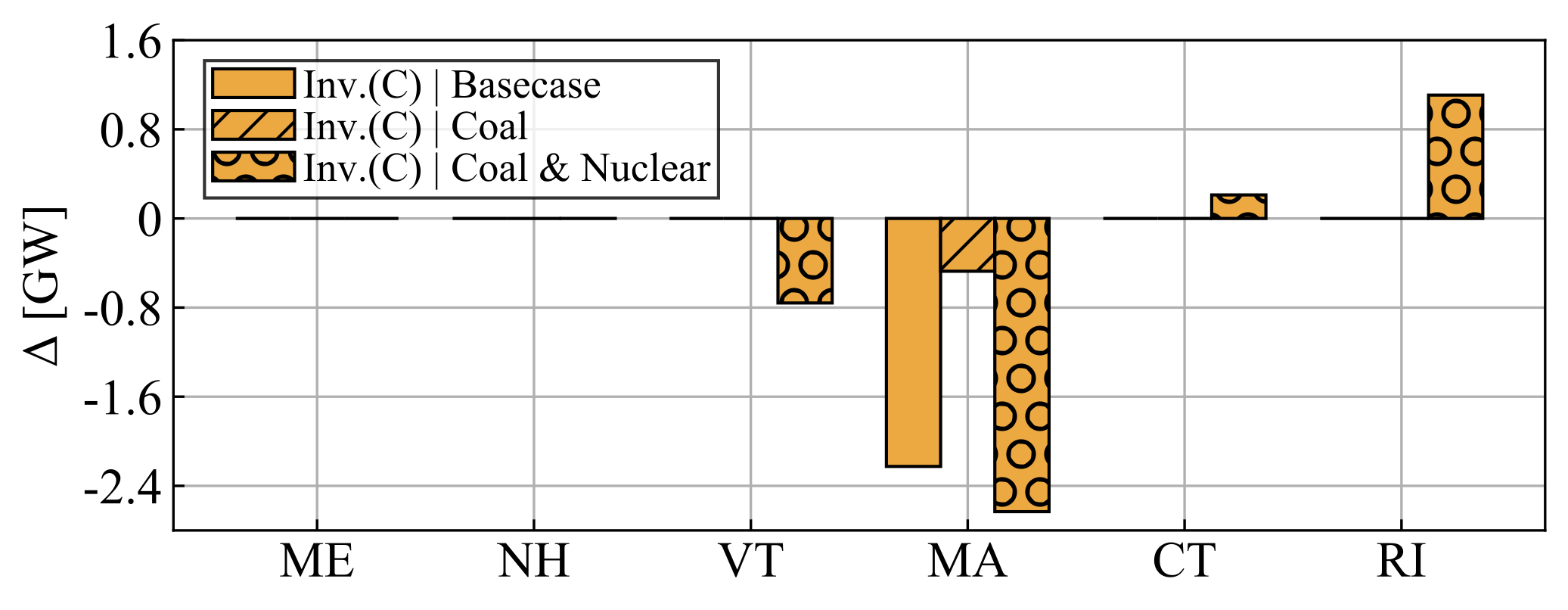}
    \vspace{-3mm}
    \caption{\small The difference of investments in controllable resources ($\Delta$, [GW]) between the CC-MPEC and the benchmark case. \vspace{1.5mm}}
    \label{fig:MPECtoBench_gpmax_CF}    
\includegraphics[scale=0.118]{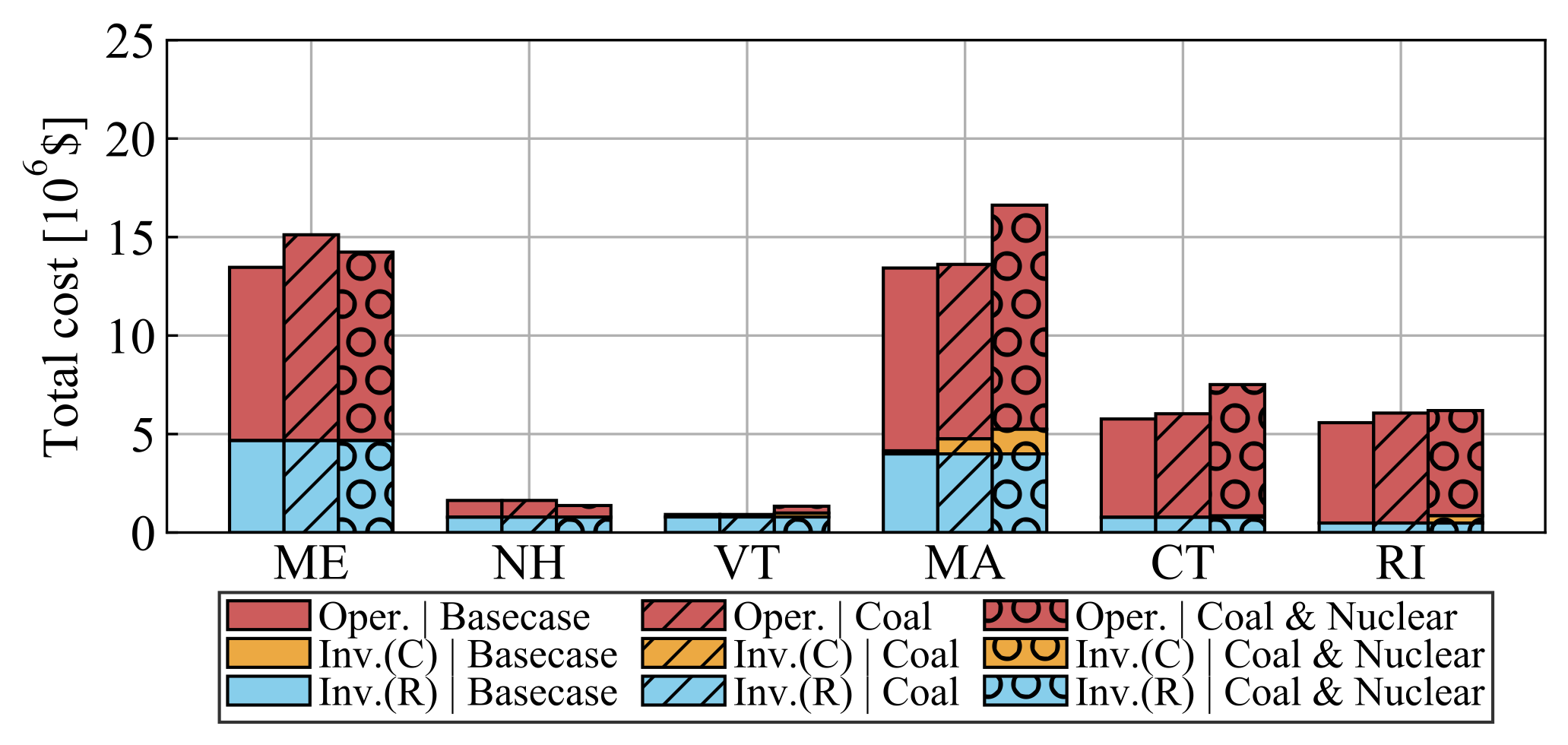}
    \vspace{-3mm}
    \caption{\small CC-MPEC: Investment and operating costs for controllable (C) and renewable (R) resources under different retirement scenarios. \vspace{1.5mm}}
    \label{fig:MPEC_CtotalF}
\end{minipage}%
\hspace{0.02\linewidth}
\begin{minipage}[b]{0.49\linewidth}
\centering
\includegraphics[scale=0.118]{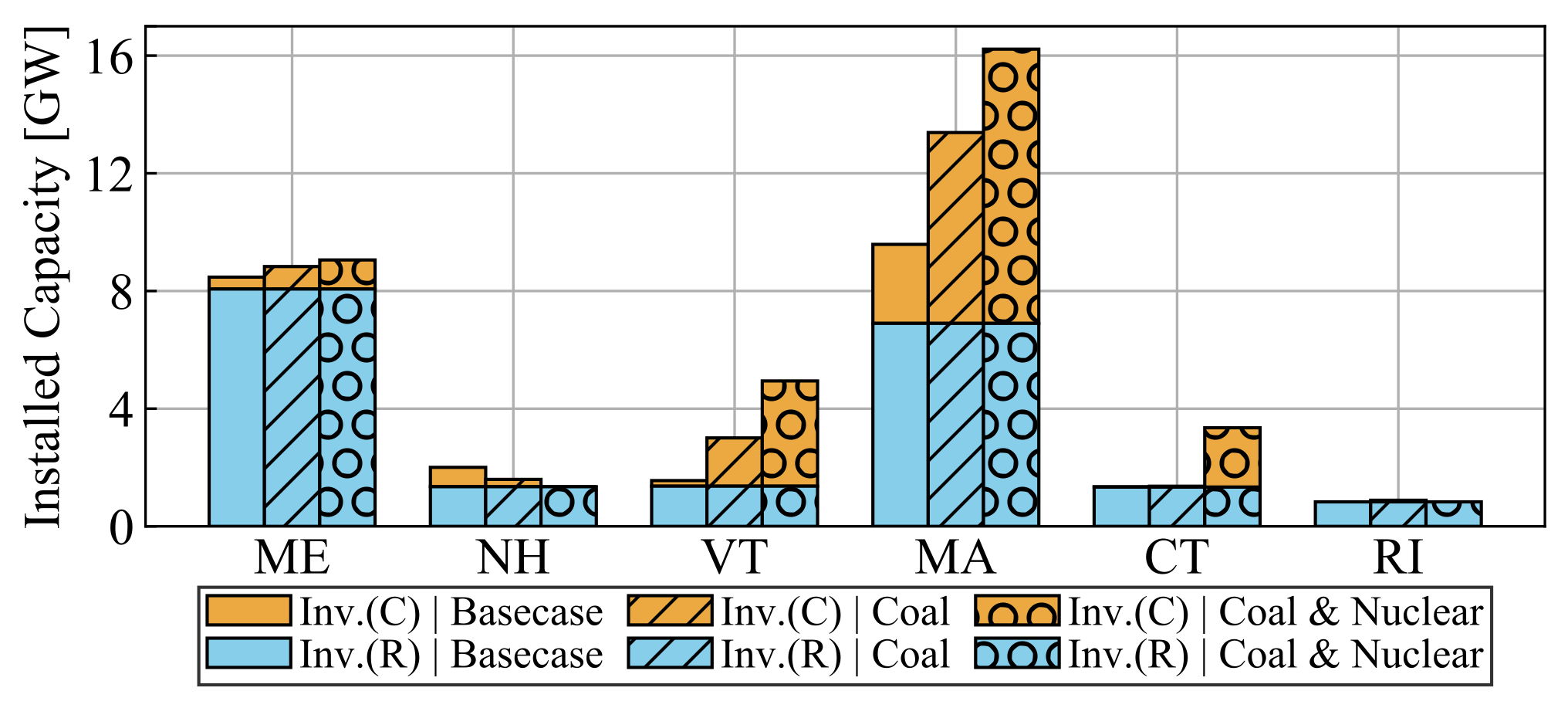}
    \vspace{-3mm}   
    \caption{\small CC-EPEC: Generation expansion decisions for controllable (C) and renewable (R) resources under different retirement scenarios. \vspace{-1.5mm}}
    \label{fig:EPEC_gpmaxF}    
\includegraphics[scale=0.118]{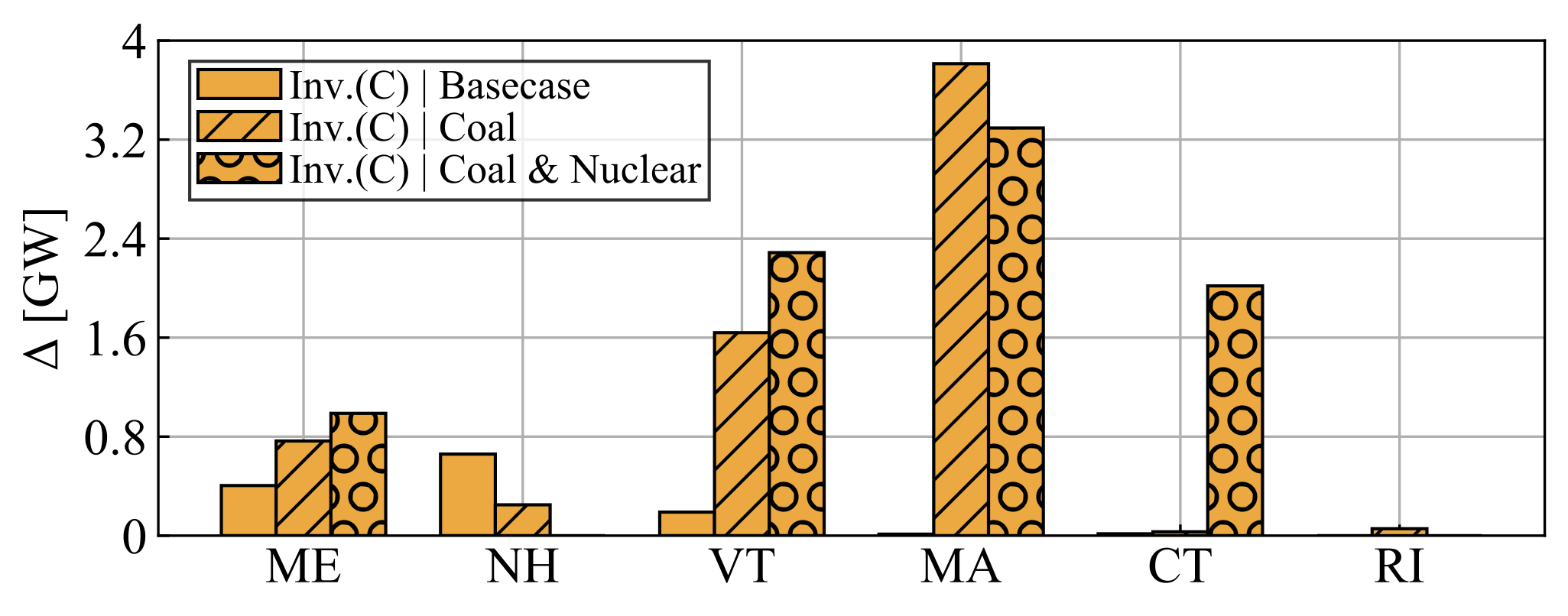}
    \vspace{-3mm}    
    \caption{\small The difference of investments in controllable resources ($\Delta$, [GW]) between the CC-EPEC and the benchmark case. \vspace{1.5mm}}
    \label{fig:EPECtoBench_gpmax_CF}   
\includegraphics[scale=0.118]{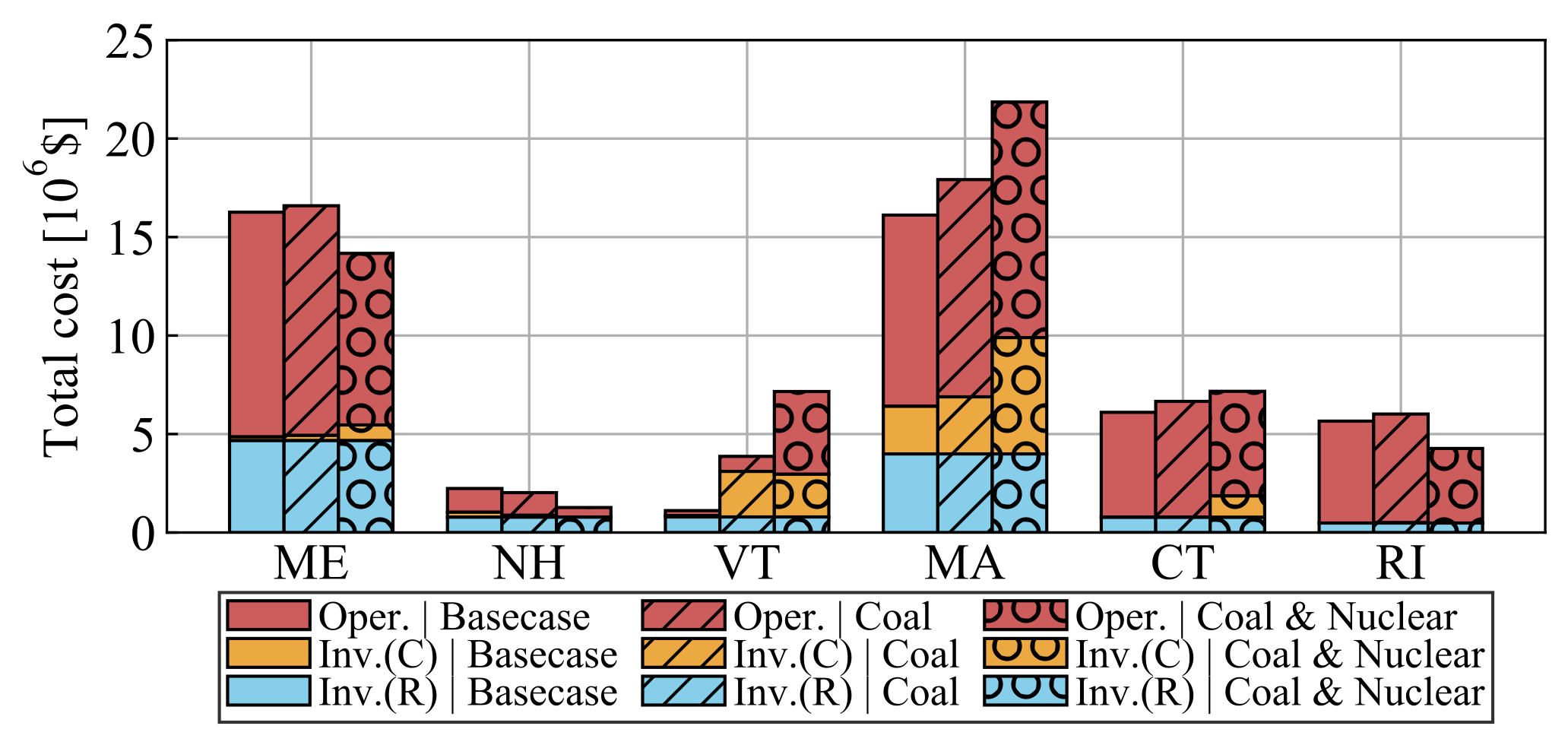}
    \vspace{-3mm}    
    \caption{\small CC-EPEC: Investment and operating costs for controllable (C) and renewable (R) resources under different retirement scenarios. \vspace{1.5mm}}
    \label{fig:EPEC_CtotalF}    
\end{minipage}%
}\vspace{-2mm}%
\end{figure*}

\vspace{-2mm}
\subsection{Case Study 1: One strategic actor}\label{subsec:CaseMPEC}
We consider one strategic actor in the ISO New England by solving the CC-MPEC in Eq.~\eqref{MPEC} for each state. 
Fig.~\ref{fig:MPEC_gpmaxF} displays investment decisions in controllable ($\sum_{i\in\hat{\cal I}^{\mathrm{C}}_s} g^{\mathrm{max}}_i$) and renewable generation resources ($\sum_{i\in\hat{\cal I}^{\mathrm{R}}_s} g^{\mathrm{max}}_i$) of the strategic actor in state $s$. In all cases, the installed renewable capacity is identical as  the RPS requirement  remains the same and installations beyond the RPS requirement are prevented by relatively high capital costs and low capacity factors, even with energy- and capacity-based renewable incentives.
On the other hand, controllable generation resources are not installed at all in ME and NH since they have a sufficient amount of existing capacity to provide reserve in all the retirement scenarios. 
However, in the cases of VT, CT and RI, the investments in controllable resources are made in the most aggressive coal \& nuclear retirement scenario.
There are two main reasons for these decisions: 1) to replenish retiring resources and 2) to export surplus generation to  neighboring states.
Notably, MA requires controllable generation installations in all retirement scenarios,  even in the no retirement case, and their capacity gradually increases as more aggressive retirement is planned.

Figure~\ref{fig:MPECtoBench_gpmax_CF} compares the investment decisions in controllable generation of the CC-MPEC and the centralized planning benchmark model.
Strategically acting ME and NH yield the same outcomes as in the benchmark. 
In the cases of VT or MA being strategic, pursing their RPS objectives requires less controllable generation installed relative to the benchmark case.
On the other hand, CT and RI need to install more controllable generation capacity than the benchmark under the coal \& nuclear retirement scenario, if they act strategically,  to achieve a lower combined capital and operating cost.
The corresponding operating and investment costs of the CC-MPECs for all retirement scenarios are compared in Fig.~\ref{fig:MPEC_CtotalF}. 

\vspace{-2mm}
\subsection{Case Study 2: Multiple strategic actors}\label{subsec:CaseEPEC}

This case study considers all states simultaneously acting strategically by solving the CC-EPEC in Eq.~\eqref{Eq:CC-EPEC}. 
Figure~\ref{fig:EPEC_gpmaxF} compares the generation expansion decisions of all six states in this case. 
Similar to the CC-MPEC case in Fig.~\ref{fig:MPEC_gpmaxF}, renewable investment decisions are the same because the RPS goals remain fixed. 
However, the installed controllable generation capacity increases relative to the CC-MPEC case.
The only exception is RI, which reduces the amount of the controllable generation installed as compared to the CC-MPEC case, due to the increased  expansion in neighboring regions and reduced transmission congestion in lines connecting RI to other regions, which makes it more cost effective to supply electricity from other states.
Finally, Fig.~\ref{fig:EPEC_CtotalF} summarizes the operating and investment costs. All costs show the same trend as in the CC-MPEC case, but the total cost for each individual region increases due to the competition. 

\vspace{-2mm}
\subsection{Computational performance}

\begin{table}[h!]
    \vspace{1mm}
    \centering
    \captionsetup{justification=centering, labelsep=period, font=footnotesize, textfont=sc}
    \caption{{\color{black}Computing times [s]}}        
    \color{black}\begin{tabular}{c|cccc}
        \hline
        \hline
        \multirow{2}{*}{Cases}&Average time&Max time&Total&Iterations\\
        &per iter&per iter&time&required\\
        \hline
        \multirow{2}{*}{Basecase}&\multirow{2}{*}{2,976}&\multirow{2}{*}{9,516}&\multirow{2}{*}{181,581}&\multirow{2}{*}{61}\\
        &&&&\\
        Coal ret.&2,476&7,365&71,802&29\\        
        Coal \&&\multirow{2}{*}{5,786}&\multirow{2}{*}{2,706}&\multirow{2}{*}{92,577}&\multirow{2}{*}{19}\\[-3pt]
        Nuclear ret.&&&&\\[+3pt]
        \hline
        \hline              
    \end{tabular}
    \label{table:comp_PH}              
    \vspace{-0mm}
\end{table}

\begin{figure}[h!]
    \vspace{-3mm}
    \centering    
    \includegraphics[width=\columnwidth]{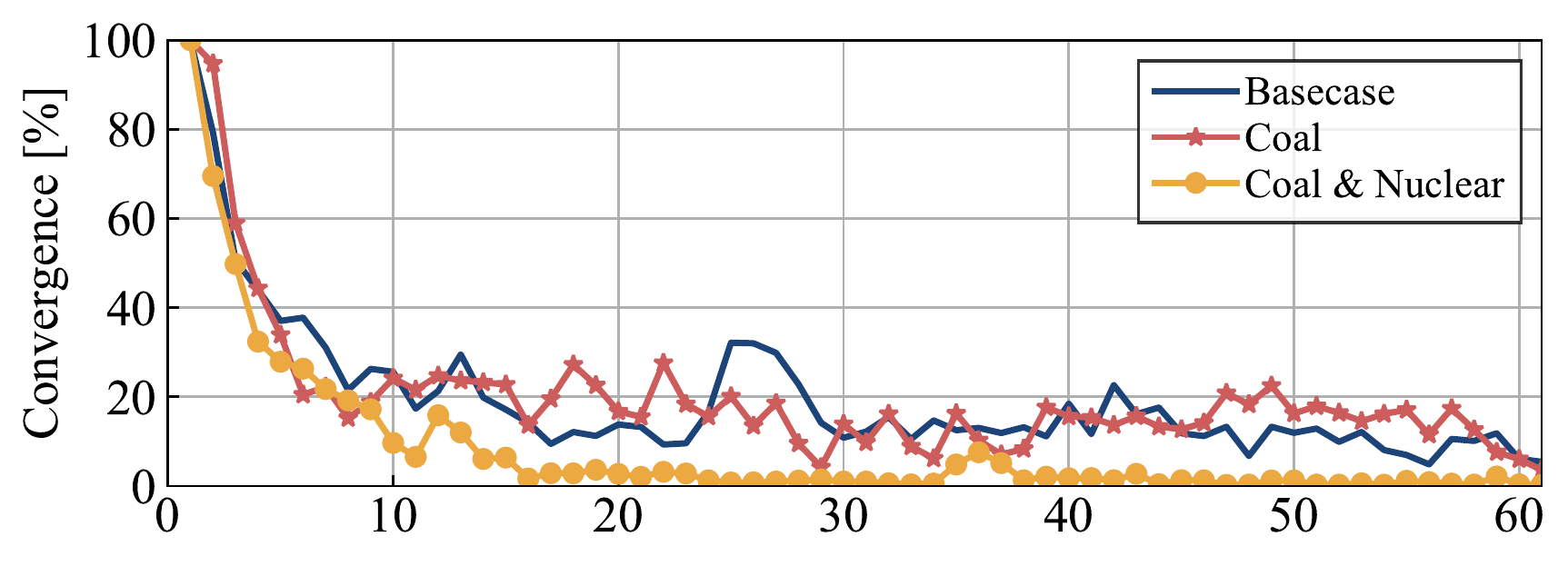}\vspace{-1mm}    
    \caption{\small Relative convergence of the proposed PH algorithm. \vspace{-3mm}}
    \label{fig:conv}    
\end{figure} 

All CC-MPEC instances  are solved within 6 hours and the longest instance was solved within 7 hours. The  CC-MPEC solution for each state was used as an initial solution (Step 2) in Algorithm 1. 
Table~\ref{table:comp_PH} summarizes computing times of the proposed PH algorithm under different retirement scenarios, where solving each augmented MPEC subproblem (Step 4-2) varies from 1 minute to 2.64 hours. Such variations in solving times often occur in the PH implementations because subproblems with solutions ($g_{ites}^{(i)}, \lambda_{ntes}^{(i)}$)  away from the consensus solution ($\mathbb{E}_s[g_{ites}^{(i)}],\mathbb{E}_s[ \lambda_{ntes}^{(i)}]$) take more time to be solved. On the other hand, subproblems with solutions in a  relative proximity to the consensus solution are typically solved faster. \cite{watson2011progressive}.
All instances for each retirement case require  32 hours on average and tens of iterations to converge as shown in Fig.~\ref{fig:conv}.

\section{Conclusion}
This paper develops a CC-EPEC model to investigate the effects of regulatory competition on achieving RPS goals in six states  in the ISO New England system. This CC-EPEC is computationally demanding and, therefore, the PH algorithm was customized to solve  the CC-EPEC efficiently. The case study enabled by the proposed model and algorithm reveals: 
\begin{enumerate} 
    \item{Compared to the uniform-policy  benchmark, strategic behavior tends to reduce the controllable generation capacity needed to support RPS goals for a single strategic actor in all but aggressive retirement scenarios}
    \item{On the other hand, the case with multiple strategic actors leads to an increased expansion of controllable generators, regardless of the retirement scenario}
    \item{Aggressive retirement scenarios strongly influence the controllable generation capacity   needed to achieve RPS goals, regardless of the number of strategic actors.}
\end{enumerate}

\bibliographystyle{IEEEtran} 
\bibliography{StrReg}

\vspace{-0mm}          
\appendix
\section{Appendix}
The benchmark model is based on the model from  \cite{munoz2014engineering}, implemented with appropriate modifications as follows:
\begin{subequations}\label{Eq:bench}
    \begin{align}        
        \begin{split}
            & \max  O^{\mathrm{B}} := \sum_{e\in{\cal E}} \Big[ \omega_e \sum_{t\in{\cal T}} \Big(\sum_{n\in{\cal N}} \hspace{-1mm}\pi_{nt}^{\mathrm{D}}D_{nte}             
             \!-\! \!\!\sum_{i\in{\cal I}\cup\hat{\cal I}}\hspace{-1.5mm} C^{\mathrm{g}}_i {g}_{ite} 
             \hspace{-10mm}
             \\&\hspace{5mm}- \hspace{-4mm}\sum_{i\in{\cal I}^{\mathrm R}\cup\hat{\cal I}^{\mathrm R}} \!\!\!\!\!\!P^{\mathrm{ET}}_{s(i)} {g}_{ite} 
             \Big) \Big]   
             -  \!\!\sum_{i\in \hat{\cal I}^{\mathrm{R}}} \!\!P^{\mathrm{CT}}_{s(i)} {g}^{\mathrm{max}}_{i}           
            - \!\!\sum_{i\in \hat{\cal I}} C^{\mathrm{inv}}_i {g}^{\mathrm{max}}_{i}
            \hspace{-10mm}
        \end{split}\label{Eq:Bench_DSO_obj}\\
        & \text{subject to:} \nonumber\\    
        & {G}_{i}^{\mathrm{min}} \le {g}_{ite} \le {G}_{i}^{\mathrm{max}} \!-\! {r}_{ite},~\forall i \in {\cal I}^{\mathrm{C}}, t \in {\cal T}, e\in{\cal E}, \label{Eq:Bench_gen_ULB} \\
        & \Gamma_{i}{g}_{i}^{\mathrm{max}} \le {g}_{ite} \le {g}_{i}^{\mathrm{max}} \!-\! {r}_{ite},~\forall i \in  \hat{\cal I}^{\mathrm{C}}, t \in {\cal T}, e\in{\cal E},\label{Eq:Bench_gen_ULB_New}\\
        \begin{split}
            &\sum_{i \in {\cal I}_n\cup\hat{\cal I}_n } \!\!\!\!\!{g}_{ite}  + \!\!\!\!\!\sum_{l \vert r(l) = n } \!\!\! f_{lte} \!-\!\!\!\!\! \sum_{l \vert o(l) = n } \!\!\! f_{lte} = D_{nte}, 
            \\& \qquad~\forall n \in {\cal N}, t\in{\cal T},e\in{\cal E},\!\!
        \end{split}\label{Bench_TSO_P_balance} \\
        & \sum_{i \in {\cal I}^{\mathrm{C}}\cup\hat{\cal I}^{\mathrm{C}}} \hspace{-1.5mm} {r}_{ite} \ge R_{te},\quad\forall t\in{\cal T}, e\in{\cal E},\\
        &f_{lte} = \frac{1}{X_l} (\theta_{o(l),t,e}-\theta_{r(l),t,e}),~\forall {l \in {\cal L}},~t\in{\cal T},e\in{\cal E},\label{Bench_TSO_DSPF}\\
        &-{F}^{\mathrm{max}}_l \le f_{lte} \le {F}^{\mathrm{max}}_l,~\forall {l \in {\cal L}},~t\in{\cal T},e\in{\cal E},\!\!\label{Bench_TSO_LineLim} \\
        & \sum_{t \in {\cal T}} \sum_{i \in {\cal I}_s^{\mathrm R}\cup \hat{\cal I}_s^{\mathrm R}} \!\!\overline{g}_{ite} \ge \kappa_s \sum_{t \in {\cal T}} \sum_{n\in{\cal N}_s} {D}_{nte},~\forall e\in{\cal E}, \label{Eq:Bench_StrReg_kappa}\\
        & \sum_{i\in\hat{\cal I}_s} C^{\mathrm{inv}}_i {g}^{\mathrm{max}}_{i}  \le  B^{\mathrm{C}}_s,\label{Eq:Bench_Cap_budget} \\
        & \sum_{t \in {\cal T}} \bigg(\! P^{\mathrm{ET}}_{s} \!\!\!\!\!\!\sum_{i\in{\cal I}_s^{\mathrm R}\cup\hat{\cal I}_s^{\mathrm R}} \!\!\!\!\!\!{g}_{ite} \bigg)\!+\! P^{\mathrm{CT}}_{s}  \sum_{i\in\hat{\cal I}_s^{\mathrm R}} \!{g}^{\mathrm{max}}_{i}   \le  B^{\mathrm{P}}_s, ~\forall e\in{\cal E}.\label{Eq:Bench_Tariff_budget}        
    \end{align}
\end{subequations}
\end{document}